\documentclass[12pt]{iopart}

\usepackage{iopams}
\usepackage{xcolor}
\usepackage{graphicx}
\usepackage[utf8]{inputenc}
\usepackage[normalem]{ulem}
\usepackage{bbm}
\usepackage{amsmath}
\usepackage{amstext}
\usepackage{cite}

\begin{document}

\def\be{\begin{equation}}
\def\ee{\end{equation}}
\def\bea{\begin{eqnarray}}
\def\eea{\end{eqnarray}}

\title{Reconstructing quantum entropy production to probe irreversibility and correlations}

\author{Stefano Gherardini$^{1,2}$, Matthias M. M\"uller$^1$, Andrea Trombettoni$^{3,4}$, Stefano Ruffo$^{4,5}$, Filippo Caruso$^1$}
\address{$^1$\mbox{Department of Physics, LENS and QSTAR, University of Florence,} via G. Sansone 1, I-50019 Sesto Fiorentino, Italy.}
\address{$^2$\mbox{Department of Information Engineering, University of Florence,} via S. Marta 3, I-50139 Florence, Italy \& \mbox{INFN},
Sezione di Firenze, Sesto Fiorentino, Italy.}
\address{$^3$\mbox{CNR-IOM DEMOCRITOS}, via Bonomea 265, I-34136 Trieste, Italy.}
\address{$^4$\mbox{SISSA}, via Bonomea 265, I-34136 Trieste, Italy \& \mbox{INFN}, Sezione di Trieste, I-34151 Trieste, Italy.}
\address{$^5$\mbox{ISC-CNR}, Via Madonna del Piano 10, I-50019 Sesto Fiorentino, Italy.}

\date{\today}

\begin{abstract}
One of the major goals of quantum thermodynamics is the characterization of irreversibility and its consequences in quantum processes. Here, we discuss how entropy production
provides a quantification of the irreversibility in open quantum systems through the quantum fluctuation theorem. We start by introducing a two-time quantum measurement scheme, in which the dynamical evolution between the measurements is described by a completely positive, trace-preserving (CPTP) quantum map (forward process).
By inverting the measurement scheme and applying the time-reversed version of the quantum map, we can study how this backward process differs from the forward one. When the CPTP map is unital, we show that the stochastic quantum entropy production is a function only of the probabilities to get the initial measurement outcomes in correspondence
of the forward and backward processes. For bipartite open quantum systems we also prove that the mean value of the stochastic quantum entropy production is sub-additive with respect to the bipartition (except for product states). Hence, we find a method to detect correlations between the subsystems. Our main result is the proposal of an efficient protocol to determine and reconstruct the characteristic functions of the stochastic entropy production for each subsystem. This procedure enables to reconstruct even others thermodynamical quantities, such as the work distribution of the composite system and the corresponding internal energy. Efficiency and possible extensions
of the protocol are also discussed. Finally, we show how our findings might be experimentally tested by exploiting the state-of-the-art trapped-ion platforms.
\end{abstract}


\maketitle

\section{Introduction}

The advent of the thermodynamics laws and their following development, from the theoretical side, and the construction of heat engines, from the technological one, drove in the 18th and 19th centuries an astonishing series of important scientific discoveries and social transformations. The crucial point was the use of heat to produce work, which corresponds to take a disordered form of energy and convert (a part of) it into a mechanical one~\cite{Groot1984}. In the last decades, a breakthrough in non-equilibrium thermodynamics was given by the Jarzynski equality~\cite{JarzynskiPRL1997}, which relates the free-energy between two equilibrium states to an exponential average of the work done on the system, over an ideally infinite number of repeated non-equilibrium experiments. This result links together free-energy differences to work measurements along an ensemble of trajectories in the phase space of the system with same energy contribution~\cite{Jarzynski2011}. The Jarzynski equality can be derived also from the Crooks fluctuation theorem~\cite{CrooksPRE1999}, which formalizes the existence of symmetry relations for the probability distribution of thermodynamic quantities during the forward and reverse transformations that the system undergoes due to external actions. Generalized versions of the Jarzynski equality for non-equilibrium steady states from Langevin dynamics and non-equilibrium systems subjected to feedback control have been proved, then, respectively in Refs.~\cite{HatanoPRL2001} and~\cite{SagawaPRL2010}. From the experimental side, the Jarzynski equality and its generalizations have been tested by a wide range of experiments, for example to determine the folding and unfolding free energies of a small RNA hairpin~\cite{CollinNAT2005}, or to prove the fundamental principle given by the information-to-heat engine, converting information into energy by means of feedback control~\cite{ToyabeNAT2010}. Even from a purely classical point of view, the notion of thermodynamics quantities such as work, heat and entropy production have been extended to the level of individual trajectories of well-defined non-equilibrium ensembles by the stochastic thermodynamics~\cite{Seifert2005,TietzPRL2006}, which has allowed for the introduction of a generalized fluctuation-dissipation theorem involving entropy production.

At the same time, the attention moved also towards the attempts to build a thermodynamic theory for quantum systems to exploit the power and the processes of quantum physics~\cite{Gemmer2004,Horodecki2013,PekolaNAT2015,Alhambra2016}. This field of research, known as quantum thermodynamics, aims at characterizing the thermodynamical aspects behind the quantum mechanical processes, defining the role of quantum coherence and measurements for such transformations~\cite{Lostaglio2015,Narasimhachar2015,LostaglioPRX2015,Kammerlander2016}. Quantum thermodynamics, moreover, provides the theoretical tools to describe and build efficient quantum heat engines~\cite{KimPRL2011,AbahPRL2012,RossnagelPRL2014}. One of the major goals of quantum thermodynamics is the definition and characterization of irreversibility in quantum processes. This could have a significant impact on technological applications for the possibility of producing  work with heat engines at high efficiency using systems where quantum fluctuations are important; in this regard, a detailed analysis about the aspects that define the work done by a quantum system can be found in~\cite{Talkner2016}. The quantum work and its distribution are generally defined by taking into account also the role of quantum measurements and, consequently, the sensitivity of the system to the interactions with the measurement apparatus~\cite{Campisi2011,HekkingPRL2013,VenkateshNJP2015,AlonsoPRL2016}. Recently a novel definition of quantum work has been proposed in~\cite{DeffnerPRE2016}, in which the work is a thermodynamic quantity depending only on the quantum system and not on the measurement apparatus.

The importance of defining the concept of irreversibility in quantum thermodynamics can be hardly overestimated, as one can appreciate by considering its classical counterpart. As well known, in classical mechanics the solutions of the dynamical equations of motion are unique and the motion along the trajectories in phase space can be inverted to retrieve all the states previously occupied by the system~\cite{Seifert2012Review}. However, the time inversion in experiments with a macroscopic number of particles cannot be practically performed. As a consequence of the information loss and of the fact that is very improbable to occupy the same state at a later time, we have to resort to a statistical description of the system. In classical thermodynamics this is the origin of the irreversibility of the system dynamics. Similarly, in quantum mechanics the dynamics of the wave function and more generally of the density matrix can be reversed in time, and it ensues the corresponding need to characterize and quantify, where possible, irreversible quantum processes~\cite{Esposito2009,CampisiRMP2011}. The typical instance is given by the thermalization of an open system, where the dissipative processes taking place due to the interaction of the system with its environment degrade the quantum nature of the system and the coherence of the quantum states~\cite{Riera2012,Gogolin2016}. Along this line, several studies have shown how to derive the quantum version of the fluctuation-dissipation theorem, both for closed~\cite{Kurchan2001,BuninNAT2011} and open quantum systems~\cite{Campisi2009,Campisi2010,Kafri2012,RasteginJSM2013,Albash2013,WatanabePRE2014,Manzano2015}. Recently, in~\cite{Aberg2016} a fully quantum fluctuation theorem have been formulated, explicitly including the reservoir exchanging energy with the system, and a control system driving its dynamics. In~\cite{Huber2008,An2015}, moreover, experimental tests of the quantum version of the Jarzynski identity~\cite{Mukamel2003,Chernyak2004,CrooksJSM2008} for work distributions are shown.

Considerable efforts have been made in measuring irreversibility, and, consequently, the stochastic entropy production in quantum thermodynamics~\cite{DeffnerPRL2011,BatalhaoPRL2015,Brunelli2016,Frenzel2016}. The ratio between the probability to observe a given quantum trajectory and its time reversal is related to the amount of heat exchanged by the quantum system with the environment~\cite{Sagawa2014}. Such knowledge leads then to experimental procedures for the measure of the heat backflow with the environment, even if the latter is not necessarily correlated with the information back-flow from the reservoir to the quantum system~\cite{Schmidt2016}. Lately, it has been experimentally proved that irreversibility in quantum non-equilibrium dynamics can be partially rectified by the presence of an \textit{intelligent observer}, identified by the well-known Maxwell's demon~\cite{KoskiPRL2014,GooldJPA2016}, which manages to assess additional microscopic informational degrees of freedom due to a proper feed-forward strategy~\cite{CamatiPRL2016}. Instead, regarding the reconstruction of the fluctuation properties of general thermodynamical quantities, in Ref.~\cite{MazzolaPRL2013,DornerPRL2013,GooldPRE2014,Peterson2016} an interferometric setting for the measurement of the characteristic function of the work distribution is introduced and proposed as the key element to properly design inference strategies~\cite{Liu2016}. This method, then, has been generalized for open quantum systems, as shown in~\cite{CampisiNJP2014,FuscoPRX2014}. In~\cite{RoncagliaPRL2014}, instead, a method for the sampling of the work distribution by means of a projective measurement at a single time is shown, motivating a novel quantum algorithm for the estimation of free energies in closed quantum systems.

In the present work we address three issues.
{\it (i)} We discuss how to relate the stochastic entropy production to the quantum fluctuation theorem, generalizing the Tasaki-Crooks theorem for open systems. This relation is obtained via the evaluation of the irreversibility of the quantum dynamics. {\it (ii)} Then, once the stochastic quantum entropy production has been defined and characterized, we introduce a protocol to reconstruct it from the measurement data, possibly with the minimum amount of resources. Here, we propose a procedure to reconstruct the stochastic entropy production of an open quantum system by performing repeated two-time measurements, at the initial and final times of the system transformation. In particular, the proposed reconstruction algorithm requires to determine the characteristic function of the probability distribution of the stochastic quantum entropy production. Indeed, by means of a parametric version of the integral quantum fluctuation theorem, we can derive the statistical moments of the entropy production. Moreover, we also prove that with this procedure the number of required measurements scales linearly with the system size. {\it (iii)} By assuming that the quantum system is bipartite, we apply the reconstruction procedure both for the two subsystems and for the composite system by performing measurements, respectively, on local and global observables. The comparison between the local and the global quantity allows us to probe the presence of correlations between the partitions of the system.

The manuscript is organized as follows. Section 2 reviews the quantum fluctuation theorem, introducing the definition of stochastic quantum entropy production. Section 3 analyzes the physical meaning of thermodynamic irreversibility by applying a two-time measurement scheme, and shows the relation between the mean entropy production and the quantum relative entropy of the system density matrix after an arbitrary transformation. The derivation in section 3 sheds light on the importance to design protocols to effectively measure the entropy production of a quantum system. In section 4 we derive the characteristic functions of the probability distributions of the stochastic entropy production within a quantum multipartite system, while in section 5 the reconstruction algorithm is introduced. We propose an experiment implementation with trapped ions in section 6. Finally, we discuss our results and conclusions in section 7.

\section{Quantum fluctuation theorem}

The fluctuations of the stochastic quantum entropy production obey the quantum fluctuation theorem. The latter can be derived by evaluating the forward and backward protocols for a non-equilibrium process, according to a two-time quantum measurement scheme~\cite{Esposito2009,CampisiRMP2011}. In this section, we introduce this two-time quantum measurement scheme and define the stochastic quantum entropy production. Then we review the derivation of the quantum fluctuation theorem.

We consider an open quantum system that undergoes a transformation in the interval $[0,\tau]$ consisting of measurement, dynamical evolution and second measurement. We call this forward process and then study also its time-reversal, which we call backward process:
\begin{eqnarray*}
\text{FORWARD}:~\rho_{0}\underbrace{\longmapsto}_{\{\Pi^{\textrm{in}}_{m}\}}\rho_{\textrm{in}}\underbrace{\longmapsto}_{\Phi}\rho_{\textrm{fin}}
\underbrace{\longmapsto}_{\{\Pi^{\textrm{fin}}_{k}\}}\rho_{\tau} \\
\text{BACKWARD}:~\widetilde{\rho}_{\tau}\underbrace{\longmapsto}_{\{\widetilde{\Pi}^{\textrm{ref}}_{k}\}}\widetilde{\rho}_{\textrm{ref}}
\underbrace{\longmapsto}_{\widetilde{\Phi}}\widetilde{\rho}_{\textrm{in}'}\underbrace{\longmapsto}_{\{\widetilde{\Pi}^{\textrm{in}}_{m}\}}\widetilde{\rho}_{0'}
\end{eqnarray*}
At time $t=0^-$ the system is prepared in a state $\rho_{0}$ and then subjected to a measurement of the observable $\mathcal{O}_{\textrm{in}} = \sum_{m}a^{\textrm{in}}_{m}\Pi^{\textrm{in}}_{m}$, where $\Pi^{\textrm{in}}_{m}\equiv|\psi_{a_{m}}\rangle\langle\psi_{a_{m}}|$ are the projector operators given in terms of the eigenvectors $|\psi_{a_{m}}\rangle$ associated to the eigenvalues $a^{\textrm{in}}_{m}$ (the $m-$th possible outcome of the first measurement). After the first measurement (at $t = 0^{+}$), the density operator describing the ensemble average of the post-measurement states becomes
\begin{equation}
\rho_{\textrm{in}} =
\sum_{m}p(a^{\textrm{in}}_{m})|\psi_{a_{m}}\rangle\langle\psi_{a_{m}}|,
\end{equation}
where $p(a^{\textrm{in}}_{m}) = \textrm{Tr}\left[\Pi^{\textrm{in}}_{m}\rho_{0}\Pi^{\textrm{in}}_{m}\right] = \langle\psi_{a_{m}}|\rho_{0}|\psi_{a_{m}}\rangle$
is the probability to obtain the measurement outcome $a^{\textrm{in}}_{m}$. Then, the system undergoes a time evolution, which we assume described by a \textit{unital} completely positive, trace-preserving (CPTP) map $\Phi:L(\mathcal{H})\rightarrow L(\mathcal{H})$, with $L(\mathcal{H})$ denoting the sets of density operators (non-negative operators with unit trace) defined on the Hilbert space $\mathcal{H}$. Quantum maps (known also as quantum channels) represent a very effective tool to describe the effects of the noisy interaction of a quantum system with its environment~\cite{Nielsen2000,Petruccione2003,Caruso14}. A CPTP map is unital if it preserves the identity operator $\mathbbm{1}$ on $\mathcal{H}$, {\em i.e.} $\Phi(\mathbbm{1}) = \mathbbm{1}$. The assumption of a unital map covers a large family of quantum physical transformations not increasing the purity of the initial states, including, among others, unitary evolutions and decoherence processes. We will briefly discuss later how the protocol presented in section 5 may be modified when the unital map hypothesis is relaxed. The time-evolved ensemble average is then denoted as
\begin{equation}
\rho_{\textrm{fin}} \equiv \Phi(\rho_{\textrm{in}})
\end{equation}
For example, in case of unitary evolution with Hamiltonian $H(t)$, the final quantum state at $t = \tau^{-}$ equals to $\rho_{\textrm{fin}} = \Phi(\rho_{\textrm{in}}) = \mathcal{U}\rho_{\textrm{in}}\mathcal{U}^{\dagger}$, where $\mathcal{U}$ is the unitary time evolution operator given by $\mathcal{U}=\mathbb{T}\exp\left(-\frac{i}{\hbar}\int^{\tau}_{0}H(t)dt\right)$, with $\mathbb{T}$ time-ordering operator. After the time evolution, at time $t = \tau^+$, a second measurement is performed on the quantum system according to the observable $\mathcal{O}_{\textrm{fin}} = \sum_{k}a^{\textrm{fin}}_{k}\Pi^{\textrm{fin}}_{k}$, where $\Pi^{\textrm{fin}}_{k} \equiv |\phi_{a_{k}}\rangle\langle\phi_{a_{k}}|$, and $a^{\textrm{fin}}_{k}$ is the $k-$th outcome of the second measurement (with eigenvectors $|\phi_{a_{k}}\rangle$). Consequently, the probability to obtain the measurement outcome $a_{k}^{\textrm{fin}}$ is $p(a^{\textrm{fin}}_{k}) = \textrm{Tr}\left[\Pi^{\textrm{fin}}_{k}\Phi(\rho_{\textrm{in}})\Pi^{\textrm{fin}}_{k}\right] =
\langle\phi_{a_{k}}|\rho_{\textrm{fin}}|\phi_{a_{k}}\rangle$. The resulting density operator, describing the ensemble average of the post-measurement states after the second measurement, is
\begin{equation}
\rho_{\tau} = \sum_{k}p(a^{\textrm{fin}}_{k})|\phi_{a_{k}}\rangle\langle\phi_{a_{k}}|.
\end{equation}
Thus, the joint probability that the events ``measure of $a^{\textrm{in}}_{m}$'' and ``measure of $a^{\textrm{fin}}_{k}$'' both occur for the forward process, denoted by $p(a^{\textrm{fin}} = a^{\textrm{fin}}_{k}, a^{\textrm{in}} = a^{\textrm{in}}_{m})$, is given by
\begin{equation}
p(a^{\textrm{fin}}_{k},a^{\textrm{in}}_{m}) = \textrm{Tr}\left[\Pi^{\textrm{fin}}_{k}\Phi(\Pi^{\textrm{in}}_{m}\rho_{0}\Pi^{\textrm{in}}_{m})\right].
\end{equation}

To study the backward process, we first have to introduce the concept of time-reversal. Time-reversal is achieved by the time-reversal operator $\Theta$ acting on $\mathcal{H}$. The latter has to be an antiunitary operator. An antiunitary operator $\Theta$ is anti-linear, {\em i.e.}
\begin{equation}
\Theta (x_1|\varphi_1\rangle + x_2|\varphi_2\rangle)= x_1^\star \Theta|\varphi_1\rangle + x_2^\star \Theta|\varphi_2\rangle
\end{equation}
for arbitrary complex coefficients $x_1$, $x_2$ and $|\varphi_1\rangle$, $|\varphi_2\rangle$ $\in$ $\mathcal{H}$, and it transforms the inner product as
$\langle \widetilde{\varphi}_1|\widetilde{\varphi}_2\rangle=\langle\varphi_2|\varphi_1\rangle$
for $|\widetilde{\varphi}_1\rangle=\Theta|\varphi_1\rangle$, and $|\widetilde{\varphi}_2\rangle=\Theta|\varphi_2\rangle$.
Antiunitary operators satisfy the relations $\Theta^{\dagger}\Theta = \Theta\Theta^{\dagger} = \mathbbm{1}$. The antiunitarity of $\Theta$ ensures the time-reversal symmetry~\cite{Sozzi2008}. We define the time-reversed density operator as $\widetilde{\rho}\equiv\Theta\rho\Theta^\dagger$, and we consider the time-reversal version of the quantum evolution operator, {\em i.e.} our unital CPTP map $\Phi$. Without loss of generality, it admits an operator-sum (or Kraus) representation: $\rho_{\textrm{fin}} = \Phi(\rho_{\textrm{in}}) = \sum_{u}E_{u}\rho_{\textrm{in}}E_{u}^{\dagger}$ with the Kraus operators $E_{u}$ being such that $\sum_{u}E_{u}^{\dagger} E_{u} = \mathbbm{1}$ (trace-preserving)~\cite{Nielsen2000,Petruccione2003,Caruso14}.
For each Kraus operator $E_{u}$ of the forward process we can define the corresponding time-reversed operator $\widetilde{E}_{u}$~\cite{CrooksPRA2008,Manzano2015}, so that the time-reversal $\widetilde{\Phi}$ for the CPTP quantum map $\Phi$ is given by
\begin{equation}\label{reversed_map}
\widetilde{\Phi}(\rho) = \sum_{u}\widetilde{E}_{u}\rho\widetilde{E}_{u}^{\dagger},
\end{equation}
where $\widetilde{E}_{u} \equiv \mathcal{A}\pi^{1/2}E^{\dagger}_{u}\pi^{-1/2}\mathcal{A}^{\dagger}$, $\pi$ is an invertible fixed point (not necessarily unique) of the quantum map, such that $\Phi(\pi) = \pi$, and $\mathcal{A}$ is an arbitrary (unitary or anti-unitary) operator. Usually, the operator $\mathcal{A}$ is chosen equal to the time-reversal operator $\Theta$. If the density operator $\pi$
is a positive definite operator, as assumed in Refs.~\cite{CrooksPRA2008,Horowitz2013}, then also the square root $\pi^{1/2}$ is positive definite and the inverse $\pi^{-1/2}$ exists and it is unique. Since our map is unital we can choose $\pi^{1/2}=\pi^{-1/2}=\mathbbm{1}$. Thus, from Eq.~(\ref{reversed_map}), we can observe that also $\widetilde{\Phi}$ is a CPTP quantum map with an operator sum-representation, such that $\sum_{u}\widetilde{E}_{u}^{\dagger}\widetilde{E}_{u} = \mathbbm{1}$.
Summarizing, we have
\begin{equation*}
\widetilde{E}_{u} = \Theta E^{\dagger}_{u}\Theta^\dagger,
\end{equation*}
so that
\begin{equation*}
\widetilde{\Phi}(\rho) = \sum_{u}\widetilde{E}_{u}\rho\widetilde{E}_{u}^{\dagger}=\Theta\left(\sum_u   E^{\dagger}_{u} \widetilde{\rho} E_{u}\right) \Theta^\dagger.
\end{equation*}
We are now in a position to define the backward process. We start by preparing the system (at time $t=\tau^+$) in the state $\widetilde{\rho}_{\tau} = \Theta\rho_{\tau}\Theta^\dagger$, and measure the observable $\widetilde{\mathcal{O}}_{\textrm{ref}} \equiv \sum_{k}a^{\textrm{ref}}_{k}\widetilde{\Pi}^{\textrm{ref}}_{k}$, with $\widetilde{\Pi}^{\textrm{ref}}_{k} = |\widetilde{\phi}_{a_{k}}\rangle\langle\widetilde{\phi}_{a_{k}}|$ and $|\widetilde{\phi}_{a_{k}}\rangle\equiv\Theta|\phi_{a_{k}}\rangle$, that is we choose this first measurement of the backward process to be the time-reversed version of the second measurement of the forward process. If we call the post-measurement ensemble average $\widetilde{\rho}_{\textrm{ref}}$, as a consequence $\widetilde{\rho}_{\tau}=\widetilde{\rho}_{\textrm{ref}}$, or equivalently $\rho_{\tau}=\rho_{\textrm{ref}}$, where the latter is called reference state. In particular, we recall that, although the quantum fluctuation theorem can be derived without imposing a specific operator for the reference state~\cite{Sagawa2014}, the latter has been chosen to be identically equal to the final density operator after the second measurement of the protocol. This choice appears to be the most natural among the possible ones to design a suitable measuring scheme of general thermodynamical quantities, consistently with the quantum fluctuation theorem. The spectral decomposition of the time-reversed reference state is given by
\begin{equation}\label{rho_ref}
\widetilde{\rho}_{\textrm{ref}} = \sum_{k}p(a^{\textrm{ref}}_{k})|\widetilde{\phi}_{a_{k}}\rangle\langle\widetilde{\phi}_{a_{k}}|,
\end{equation}
where
\begin{equation}
p(a^{\textrm{ref}}_{k}) = \textrm{Tr}[\widetilde{\Pi}^{\textrm{ref}}_{k}\widetilde{\rho}_{\tau}\widetilde{\Pi}^{\textrm{ref}}_{k}] = \langle\widetilde{\phi}_{a_{k}}|\widetilde{\rho}_{\tau}|\widetilde{\phi}_{a_{k}}\rangle
\end{equation}
is the probability to get the measurement outcome $a_{k}^{\textrm{ref}}$. The reference state undergoes the time-reversal dynamical evolution, mapping it onto the initial state of the backward process $\widetilde{\rho}_{\textrm{in}'}=\widetilde{\Phi}(\widetilde{\rho}_{\textrm{ref}})$.
At $t = 0^{+}$ the density operator $\widetilde{\rho}_{\textrm{in}'} = \widetilde{\Phi}(\widetilde{\rho}_{\textrm{ref}})$ is subject to the second projective measurement of the backward process, whose observable is given by $\widetilde{\mathcal{O}}_{\textrm{in}} = \sum_{m}a^{\textrm{in}}_{m}\widetilde{\Pi}^{\textrm{in}}_{m}$, with $\widetilde{\Pi}^{\textrm{in}}_{m} = |\widetilde{\psi}_{a_{m}}\rangle\langle\widetilde{\psi}_{a_{m}}|$, and $|\widetilde{\psi}_{a_{m}}\rangle\equiv\Theta|\psi_{a_{m}}\rangle$. As a result, the probability to obtain the outcome $a^{\textrm{in}}_{m}$ is
$p(a^{\textrm{in}}_{m}) = \textrm{Tr}[\widetilde{\Pi}^{\textrm{in}}_{m}\widetilde{\Phi}(\widetilde{\rho}_{\textrm{ref}})\widetilde{\Pi}^{\textrm{in}}_{m}] =
\langle\widetilde{\psi}_{a_{m}}|\widetilde{\rho}_{\textrm{in}'}|\widetilde{\psi}_{a_{m}}\rangle$, while the joint probability $p(a^{\textrm{in}}_{m}, a^{\textrm{ref}}_{k})$ is given by
\begin{equation}
p(a^{\textrm{in}}_{m}, a^{\textrm{ref}}_{k}) = \textrm{Tr}[\widetilde{\Pi}^{\textrm{in}}_{m}\widetilde{\Phi}(\widetilde{\Pi}^{\textrm{ref}}_{k}\widetilde{\rho}_{\tau}
\widetilde{\Pi}^{\textrm{ref}}_{k})].
\end{equation}
The final state of the backward process is instead $\widetilde{\rho}_{0'}=\sum_{m} p(a^{\textrm{in}}_{m}) \widetilde{\Pi}^{\textrm{in}}_{m}$. Let us observe again that the main difference of the two-time measurement protocol that we have introduced here, compared to the scheme in Ref.~\cite{Sagawa2014}, is to perform the $2$nd and $1$st measurement of the backward protocol, respectively, on the same basis of the $1$st and $2$nd measurement of the forward process after a time-reversal transformation.

The irreversibility of the two-time measurement scheme can be analyzed by studying the stochastic quantum entropy production $\sigma$ defined as:
\begin{equation}\label{general_sigma}
\sigma(a^{\textrm{fin}}_{k},a^{\textrm{in}}_{m}) \equiv \ln\left[\frac{p(a^{\textrm{fin}}_{k}, a^{\textrm{in}}_{m})}{p(a^{\textrm{in}}_{m}, a^{\textrm{ref}}_{k})}\right]
= \ln\left[\frac{p(a^{\textrm{fin}}_{k}|a^{\textrm{in}}_{m})p(a^{\textrm{in}}_{m})}
{p(a^{\textrm{in}}_{m}|a^{\textrm{ref}}_{k})p(a^{\textrm{ref}}_{k})}\right],
\end{equation}
where $p(a^{\textrm{fin}}_{k}|a^{\textrm{in}}_{m})$ and $p(a^{\textrm{in}}_{m}|a^{\textrm{ref}}_{k})$ are the conditional probabilities of measuring, respectively, the outcomes $a^{\textrm{fin}}_{k}$ and $a^{\textrm{in}}_{m}$, conditioned on having first measured $a^{\textrm{in}}_{m}$ and $a^{\textrm{ref}}_{k}$. Its mean value
\begin{equation}\label{general_mean_sigma}
\langle\sigma\rangle = \sum_{k,m}p(a_k^{\textrm{fin}},a_m^{\textrm{in}})\ln\left[\frac{p(a_k^{\textrm{fin}},a_m^{\textrm{in}})}{p(a_k^{\textrm{in}}, a_m^{\textrm{ref}})}\right]
\end{equation}
corresponds to the classical relative entropy (or Kullback-Leibler divergence) between the joint probabilities $p(a^{\textrm{fin}}, a^{\textrm{in}})$ and $p(a^{\textrm{in}}, a^{\textrm{ref}})$, respectively, of the forward and backward processes~\cite{Cover2006,Umegaki}. The Kullback-Leibler divergence is always non-negative and as a consequence
\begin{equation}
\langle\sigma\rangle\geq 0.
\end{equation}
As a matter of fact, $\langle\sigma\rangle$ can be considered as the amount of additional information that is required to achieve the backward process, once the quantum system has reached the final state $\rho_{\tau}$. Moreover, $\langle\sigma\rangle = 0$ if and only if $p(a^{\textrm{fin}}_{k}, a^{\textrm{in}}_{m}) = p(a^{\textrm{in}}_{m}, a^{\textrm{ref}}_{k})$, {\em i.e.} if and only if $\sigma = 0$. To summarize, the transformation of the system state from time $t=0^{-}$ to $t=\tau^{+}$ is then defined to be thermodynamically irreversible if $\langle\sigma\rangle > 0$. If, instead, all the fluctuations of $\sigma$ shrink around $\langle\sigma\rangle \simeq 0$ the system comes closer and closer to a reversible one. We observe that a system transformation may be thermodynamically irreversible also if the system undergoes unitary evolutions with the corresponding irreversibility contributions due to applied quantum measurements. Also the measurements back-actions, indeed, lead to energy fluctuations of the quantum system, as recently quantified in Ref.~\cite{Elouard2016}. In case there is no evolution (identity map) and the two measurement operators are the same, then the transformation becomes reversible. We can now state the following theorem: \\ \\
\textbf{Theorem~1}: \textit{Given the two-time measurement protocol described above and an open quantum system dynamics described by a unital CPTP quantum map} $\Phi$, \textit{it can be stated that:}
\begin{equation}\label{equality_cond_prob}
p(a^{\textrm{fin}}_{k}|a^{\textrm{in}}_{m}) = p(a^{\textrm{in}}_{m}|a^{\textrm{ref}}_{k}).
\end{equation}
The proof of Theorem~1 can be found in \ref{appendix_a}. \\

Throughout this article we assume that $\Phi$ is unital and this property of the map guarantees the validity of Theorem~1. Note, however, that Refs.~\cite{Horowitz2013,Manzano2015} present a fluctuation theorem for slightly more general maps, that however violate Eq.~(\ref{equality_cond_prob}).

As a consequence of Theorem~1 we obtain:
\begin{equation}\label{sigma}
\sigma(a^{\textrm{fin}}_{k},a^{\textrm{in}}_{m}) = \ln\left[\frac{p(a^{\textrm{in}}_{m})}{p(a^{\textrm{ref}}_{k})}\right] =\ln\left[\frac{\langle\psi_{a_{m}}|\rho_{0}|\psi_{a_{m}}\rangle}
{\langle\widetilde{\phi}_{a_{k}}|\widetilde{\rho}_{\tau}|\widetilde{\phi}_{a_{k}}\rangle}\right].
\end{equation}
providing a general expression of the quantum fluctuation theorem for the described two-time quantum measurement scheme. Let us introduce, now, the entropy production $\widetilde{\sigma}$ for the backward processes, {\em i.e.}
$$
\widetilde{\sigma}(a^{\textrm{in}}_{m},a^{\textrm{ref}}_{k})\equiv\ln\left[\frac{p(a^{\textrm{in}}_{m}, a^{\textrm{ref}}_{k})}{p(a^{\textrm{fin}}_{k}, a^{\textrm{in}}_{m})}\right] = \ln\left[\frac{p(a^{\textrm{ref}}_{k})}{p(a^{\textrm{in}}_{m})}\right],
$$
where the second identity is valid only in case we can apply the results deriving from Theorem~1. Hence, if we define $\textrm{Prob}(\sigma)$ and $\textrm{Prob}(\widetilde{\sigma})$ as the probability distributions of the stochastic entropy production, respectively, for the forward and the backward processes, then it can be shown (see e.g. Ref.~\cite{Sagawa2014}) that
\begin{equation}\label{qft}
\frac{\textrm{Prob}(\widetilde{\sigma} = -\Gamma)}{\textrm{Prob}(\sigma = \Gamma)} = e^{-\Gamma},
\end{equation}
where $\Gamma$ belongs to the set of values that can be assumed by the stochastic quantum entropy production $\sigma$. Eq.~(\ref{qft}) is usually called quantum fluctuation theorem. By summing over $\Gamma$, we recover the \textit{integral quantum fluctuation theorem}, or quantum Jarzynski equality, $\langle e^{-\sigma}\rangle = 1$, as shown e.g. in Refs.~\cite{Kurchan2001,Sagawa2014}. The role of the integral fluctuation theorem in deriving the probability distribution $\textrm{Prob}(\sigma)$ of the stochastic entropy production for an open quantum system is analyzed in the following sections.

\section{Mean entropy production vs quantum relative entropy}
\label{new_sec_3}

In this section, we discuss the irreversibility of the two-time measurement scheme for an open quantum system $\mathcal{S}$ in interaction with the environment $\mathcal{E}$ (described by a unital CPTP map), deriving an inequality (Theorem 2) for the entropy growth. Following Ref.~\cite{Sagawa2014}, the essential ingredient is the non-negativity of the quantum relative entropy and its relation to the stochastic quantum entropy production. As a generalization of the Kullback-Leibler information~\cite{Umegaki}, the quantum relative entropy between two arbitrary density operators $\nu$ and $\mu$ is defined as $S(\nu\parallel\mu)\equiv\textrm{Tr}[\nu\ln\nu] - \textrm{Tr}[\nu\ln\mu]$. The Klein inequality states that the quantum relative entropy is a non-negative quantity~\cite{Vedral2002}, {\em i.e.} $S(\nu\parallel\mu)\geq 0$, where the equality holds if and only if $\nu = \mu$ - see {\em e.g.}~\cite{Sagawa2014}. In particular, in the following theorem we will show the relation between the quantum relative entropy of the system density matrix at the final time of the transformation and the stochastic quantum entropy production for unital CPTP quantum maps: \\ \\
\textbf{Theorem~2}: \textit{Given the two-time measurement protocol described above and an open quantum system dynamics described by a unital CPTP quantum map} $\Phi$, \textit{the quantum relative entropy} $S(\rho_{\textrm{fin}}\parallel\rho_{\tau})$ \textit{fulfills the inequality}
\begin{equation}\label{eq:entropy-positivity}
0\leq S(\rho_{\textrm{fin}}\parallel\rho_{\tau})\leq\langle\sigma\rangle,
\end{equation}
\textit{where the equality} $S(\rho_{\textrm{fin}}\parallel\rho_{\tau}) = 0$ \textit{holds if and only if} $\rho_{\textrm{fin}} = \rho_{\tau}$. \textit{Then, for} $[\mathcal{O}_{\textrm{fin}},\rho_{\textrm{fin}}]=0$ \textit{one has} $\langle\sigma\rangle = S(\rho_{\tau}) - S(\rho_{\textrm{in}})$, \textit{so that}
\begin{equation}\label{eq:entropy-of-map}
0= S(\rho_{\textrm{fin}}\parallel\rho_{\tau})\leq\langle\sigma\rangle=S(\rho_{\textrm{fin}})-S(\rho_{\textrm{in}}),
\end{equation}
\textit{where $S(\cdot)$ denotes the von Neumann entropy of $(\cdot)$. Finally,} $S(\rho_{\textrm{fin}}\parallel\rho_{\tau}) = \langle\sigma\rangle$ \textit{if} $\mathcal{S}$ \textit{is a closed quantum system following a unitary evolution}. A proof of Theorem~2 is in \ref{appendix_b}. \\

While Eq.~(\ref{eq:entropy-positivity}) is more general and includes the irreversibility contributions of both the map $\Phi$ and the final measurement, in Eq.~(\ref{eq:entropy-of-map}) due to a special choice of the observable of the second measurement we obtain $\rho_{\textrm{fin}}=\rho_{\tau}$ and, thus, the quantum relative entropy vanishes while the stochastic quantum entropy production contains the irreversibility contribution only from the map. This contribution is given by the difference between the von Neumann entropy of the final state $S(\rho_{\textrm{fin}})$ and the initial one $S(\rho_{\textrm{in}})$\footnote{Let us assume that the initial density matrix $\rho_{\textrm{in}}$ is a Gibbs thermal state at inverse temperature $\beta$, {\em i.e.} $\rho_{\textrm{in}}\equiv e^{\beta\left[F(0)\mathbbm{1}_\mathcal{S}- H(0)\right]}$, where $F(0) \equiv -\beta^{-1}\ln\left\{\textrm{Tr}[e^{-\beta H(t = 0)}]\right\}$ and $H(0)$ are, respectively, equal to the Helmholtz free-energy and the system Hamiltonian at time $t=0$. Accordingly, the von Neumann entropy $S(\rho_{\textrm{in}})$ equals the thermodynamic entropy at $t = 0$, {\em i.e.} $S(\rho_{\textrm{in}}) = \beta(\langle H(0)\rangle - F(0))$, where $\langle H(0)\rangle \equiv \textrm{Tr}[\rho_{\textrm{in}}H(0)]$ is the average energy of the system in the canonical distribution. More generally, we can state that given an arbitrary initial density matrix $\rho_{\textrm{in}}$ the thermodynamic entropy $\beta(\langle H(0)\rangle - F(0))$ represents the upper-bound value for the von Neumann entropy $S(\rho_{\textrm{in}})$, whose maximum value is reached only in the canonical distribution. To prove this, it is sufficient to consider $S(\rho_{\textrm{in}}\parallel e^{\beta\left(F(0)\mathbbm{1}_\mathcal{S} - H(0)\right)}) = \beta\left(F(0) - \langle H(0)\rangle\right) - S(\rho_{\textrm{in}})$, from which, from the positivity of the quantum relative entropy, one has $S(\rho_{\textrm{in}})\leq\beta(\langle H(0)\rangle - F(0))$.}.

\subsection{Physical considerations}
To summarize the previous results, in case the environment $\mathcal{E}$ is not thermal,
the stochastic quantum entropy production represents a very general measurable thermodynamic quantity, encoding information about the interaction between the system and the environment also in a fully quantum regime. Therefore, its reconstruction becomes relevant, not only for the fact that we cannot longer adopt energy measurements on $\mathcal{S}$ to infer $\sigma$ and its fluctuation properties, but also because in this way we could manage to measure the mean heat flux exchanged by the partitions of $\mathcal{S}$ in case it is a multipartite quantum system, as shown in the following sections.

\subsection{Recovering the second law of thermodynamics}
\label{rec_2nd_law}

After stating the results of Theorem 2, and, accordingly, having discussed the irreversibility contributions coming from unital quantum CPTP maps, the following question naturally emerges about the connection between the Theorem 2, valid also for a quantum dynamics at $T=0$, and the second law of thermodynamics, given the fact
that in Theorem 2 there is no specific reference to thermal states. Therefore we think that it is useful in this Section to address the question:
How the inequality (\ref{eq:entropy-positivity}) for the entropy growth can be connected to the conventional second law of thermodynamics given in terms of the energetic quantities of the quantum system?

In a fully quantum regime, following~\cite{Alicki1979,LorenzoPRL2015}, the internal energy of a quantum system $\mathcal{S}$ is given by the relation $\textrm{Tr}[\rho(t)H(t)]\equiv
\textrm{Tr}[\rho H](t)$, where $H(t)$ is the (time-dependent) Hamiltonian of the system. Accordingly, an infinitesimal change of the internal energy
during the infinitesimal interval $[t,t+\delta t]$ will be $\delta \textrm{Tr}[\rho H](t)\equiv\textrm{Tr}[\rho(t+\delta t)H(t+\delta t)] - \textrm{Tr}[\rho(t)H(t)]$. The latter, then, can be recast into the following equation, representing the first law of thermodynamics for the quantum system:
\begin{equation}\label{eq:diff_1stlaw}
\delta \textrm{Tr}[\rho H](t) = \textrm{Tr}[\rho(t)\delta H(t)] + \textrm{Tr}[\delta\rho(t)H(t)],
\end{equation}
where $\delta H(t)\equiv H(t+\delta t)-H(t)$ and $\delta\rho(t)\equiv\rho(t+\delta t)-\rho(t)$. The quantity $\textrm{Tr}[\rho(t)\delta H(t)]$ is the infinitesimal mean work $\delta\langle\mathrm{W}\rangle(t)$ done by the system in the time interval $[t,t+\delta t]$, while $\textrm{Tr}[\delta\rho(t)H(t)]$ denotes the infinitesimal mean heat flux $\delta\langle\mathrm{Q}\rangle(t)$, which is identically equal to zero if the quantum system dynamics is unitary. Thus, the mean work done the system is $\langle W\rangle = \textrm{Tr}[\rho_{\textrm{fin}}H(\tau)] - \textrm{Tr}[\rho_{\textrm{in}}H(0)]$, while the mean heat flux $\langle\mathrm{Q}\rangle$, for a time-independent Hamiltonian and a finite value change of the internal energy of the quantum system during the protocol, equals to
$
\langle\mathrm{Q}\rangle = \textrm{Tr}[\rho_{\textrm{fin}}H] - \textrm{Tr}[\rho_{\textrm{in}}H] = \textrm{Tr}\left[(\Phi -
\mathbb{I})[\rho_{\textrm{in}}]H\right],
$
where $\mathbb{I}$ is the identity map acting on the sets of the density operators within the Hilbert space of $\mathcal{S}$.

From the results of Theorems 1 and 2, valid under the hypothesis that the quantum CPTP map of the quantum system is unital, one has that
\begin{equation}\label{sigma_theorem2}
\langle\sigma\rangle = - {\rm Tr}[\rho_{\textrm{fin}}\ln\rho_{\tau}] - S(\rho_{\textrm{in}}).
\end{equation}
In order to recover the second law of thermodynamics as a relation between the mean work $\langle\mathrm{W}\rangle$ and
the Helmholtz free-energy difference $\Delta F \equiv F(\tau) - F(0)$, we need to quantify the deviations of
${\rm Tr}[\rho_{\textrm{fin}}\ln\rho_{\tau}]$ from the corresponding value
in a thermal state $\rho_{\tau}^{{\rm th}}$, defined to be the thermal state
for the Hamiltonian $H$ at time $\tau$. In formulas,
$\rho_{\tau}^{{\rm th}} \equiv e^{\beta\left[F(\tau)\mathbbm{1}_\mathcal{S} - H(\tau)\right]}$ at time $t = \tau$. The derivation is done in
\ref{appendix_second_law}, and the result is
\begin{equation}\label{new_equation_quantum_entropy}
S(\rho_{\textrm{fin}}\parallel\rho_{\tau}) + {\rm Tr}[\rho_{\textrm{fin}}\ln\rho_{\tau}] = S(\rho_{\textrm{fin}}\parallel\rho_{\tau}^{{\rm th}}) + {\rm Tr}[\rho_{\textrm{fin}}\ln\rho_{\tau}^{{\rm th}}],
\end{equation}
so that
\begin{equation*}
{\rm Tr}[\rho_{\textrm{fin}}\ln\rho_{\tau}] = {\rm Tr}[\rho_{\textrm{fin}}\ln\rho_{\tau}^{{\rm th}}] + S(\rho_{\textrm{fin}}\parallel\rho_{\tau}^{{\rm th}})
- S(\rho_{\textrm{fin}}\parallel\rho_{\tau}).
\end{equation*}
Accordingly, by substituting $\rho_{\textrm{in}} \equiv e^{\beta\left[F(0)\mathbbm{1}_\mathcal{S} - H(0)\right]}$ and $\ln\rho_{\tau}^{{\rm th}} = \beta(F(\tau) - H(\tau))$ in Eq. (\ref{sigma_theorem2}), one has that
\begin{equation*}
\langle\sigma\rangle - S(\rho_{\textrm{fin}}\parallel\rho_{\tau}) = \beta\left(\langle\mathrm{W}\rangle - \Delta F\right) - S(\rho_{\textrm{fin}}\parallel\rho_{\tau}^{{\rm th}}),
\end{equation*}
where $\langle\sigma\rangle - S(\rho_{\textrm{fin}}\parallel\rho_{\tau})\geq 0$, since $0\leq S(\rho_{\textrm{fin}}\parallel\rho_{\tau})\leq\langle\sigma\rangle$. Finally, observing that $S(\rho_{\textrm{fin}}\parallel\rho_{\tau}^{{\rm th}})\geq 0$ being a quantum relative entropy, we recover the conventional second law of thermodynamics
\begin{equation}
\langle\mathrm{W}\rangle\geq\Delta F.
\end{equation}

The validity of the second law of thermodynamics has been proved just by exploiting the non-negativity of the quantum relative entropy and the results from Theorems 1 and 2. However, to avoid possible misunderstadigs,
let us clarify that a unital quantum process cannot in general describe the mapping between two Gibbs thermal states, and, thus, neither a thermalization process for $\mathcal{S}$. Accordingly, the density operator $\rho_{\tau}$ will not be physically equal to the corresponding thermal state $\rho_{\tau}^{{\rm th}}$, and $\langle\sigma\rangle$ is not linearly proportional (with $\beta$ as proportionality constant) to the internal energy of $\mathcal{S}$, i.e. to the mean heat flux $\langle\mathrm{Q}\rangle$. One can see
this taking Eq. (\ref{sigma_theorem2}) and substituting
in $\rho_{\textrm{in}}$ the thermal state $e^{\beta\left[F(0)\mathbbm{1}_\mathcal{S} - H(0)\right]}$: being $\rho_{\tau}$ a mixed but not thermal state, necessarily $\langle\sigma\rangle \neq \beta\left(\langle\mathrm{W}\rangle - \Delta F\right)$.
\\

\section{Stochastic quantum entropy production for open bipartite systems}

In this section, our intent is to define and, then, reconstruct the fluctuation profile of the stochastic quantum entropy production $\sigma$ for an open multipartite system (for simplicity we will analyze in detail a bipartite system), so as to characterize the irreversibility of the system dynamics after an arbitrary transformation. At the same time, we will also study the role played by the performance of measurements both on local and global observables for the characterization of $\textrm{Prob}(\sigma)$ in a many-body context, and evaluate the efficiency of reconstruction in both cases. In particular, as shown by the numerical examples, by comparing the mean stochastic entropy productions $\langle\sigma\rangle$ obtained by local measurements on partitions of the composite system and measurements on its global observables, we are able to detect (quantum and classical) correlations between the subsystems, which have been caused by the system dynamics.

To this end, let us assume that the open quantum system $\mathcal{S}$ is composed of two distinct subsystems ($A$ and $B$), which are mutually interacting, and we denote by $A-B$ the composite system $\mathcal{S}$. However, all the presented results can be in principle generalized to an arbitrary number of subsystems. As before, the initial and final density operators of the composite system are arbitrary (not necessarily equilibrium) quantum states, and the dynamics of the composite system is described by a unital CPTP quantum map. The two-time measurement scheme on $A - B$ is implemented by performing the measurements locally on $A$ and $B$ and we assume, moreover, that the measurement processes at the beginning and at the end of the protocol are independent. Since the local measurement on $A$ commutes with the local measurement on $B$, the two measurements can be performed simultaneously. This allows us to consider the stochastic entropy production for the composite system by considering the correlations between the measurement outcomes of the two local observables. Alternatively, by disregarding these correlations, we can consider separately the stochastic entropy production of each subsystem.

The composite system $A - B$ is defined on the finite-dimensional Hilbert space $\mathcal{H}_{A-B}\equiv\mathcal{H}_{A}\otimes\mathcal{H}_{B}$ (with $\mathcal{H}_A$ and $\mathcal{H}_B$ the Hilbert spaces of system $A$ and $B$, respectively), and its dynamics is governed by the following
time-dependent Hamiltonian
\begin{equation}
H(t) = H_{A}(t)\otimes\mathbbm{1}_{B} + \mathbbm{1}_{A}\otimes H_{B}(t) + H_{A-B}(t).
\end{equation}
$\mathbbm{1}_{A}$ and $\mathbbm{1}_{B}$ are the identity operators acting, respectively, on the Hilbert spaces of the systems $A$ and $B$, while $H_{A}$ is the Hamiltonian of $A$, $H_{B}$ the Hamiltonian of system $B$, and $H_{A-B}$ is the interaction term. We denote the initial density operator of the composite quantum system $A - B$ by $\rho_{0}$ (before the first measurement), which we assume to be a product state, then the ensemble average after the first measurement (at $t = 0^{+}$) is given by the density operator $\rho_{\textrm{in}}$, which can be written as:
\begin{equation}\label{rho_in}
\rho_{\textrm{in}}=\rho_{A,\textrm{in}}\otimes\rho_{B,\textrm{in}},
\end{equation}
where
\begin{equation}
\begin{cases}
\rho_{A,\textrm{in}}=\sum_{m}p(a_{m}^{\textrm{in}})\Pi^{\textrm{in}}_{A,m} \\
\rho_{B,\textrm{in}}=\sum_{h}p(b_{h}^{\textrm{in}})\Pi^{\textrm{in}}_{B,h} \\
\end{cases}
\end{equation}
are the reduced density operators for the subsystems $A$ and $B$, respectively. The projectors
$\Pi^{\textrm{in}}_{A,m} \equiv |\psi_{a_{m}}\rangle\langle\psi_{a_{m}}|$ and $\Pi^{\textrm{in}}_{B,h}\equiv|\psi_{b_{h}}\rangle\langle\psi_{b_{h}}|$ are the projectors onto the respective eigenstates of the local measurement operators for the subsystems $A$ and $B$:
the observables $\mathcal{O}^{\textrm{in}}_{A}=\sum_m a_m^{\textrm{in}}\Pi^{\textrm{in}}_{A,m}$ on system $A$ and $\mathcal{O}^{\textrm{in}}_{B}=\sum_h b_h^{\textrm{in}}\Pi^{\textrm{in}}_{B,h}$ on system $B$, with possible measurement outcomes $\{a^{\textrm{in}}_{m}\}$ and $\{b^{\textrm{in}}_{h}\}$, upon measurement of $\rho_0$. After the measurement, the composite system $A - B$ undergoes a time evolution up to the time instant $t = \tau^{-}$, described by the unital CPTP quantum map $\Phi$, such that $\rho_{\textrm{fin}}=\Phi(\rho_{\textrm{in}})$. Then, a second measurement is performed on both systems, measuring the observables $\mathcal{O}^{\textrm{fin}}_{A}=\sum_k a_k^{\textrm{fin}}\Pi^{\textrm{fin}}_{A,k}$ on system $A$ and $\mathcal{O}^{\textrm{fin}}_{B}=\sum_l b_l^{\textrm{fin}}\Pi^{\textrm{fin}}_{B,l}$ on system $B$, where $\{a_k^{\textrm{fin}}\}$ and $\{b_l^{\textrm{fin}}\}$ are the eigenvalues of the observables, and the projector $\Pi^{\textrm{fin}}_{A,k}\equiv|\phi_{a_{k}}\rangle\langle\phi_{a_{k}}|$ and $\Pi^{\textrm{fin}}_{B,l}\equiv|\phi_{b_{l}}\rangle\langle\phi_{b_{l}}|$ are given by the eigenstates $|\phi_{a_{k}}\rangle$ and $|\phi_{b_{l}}\rangle$, respectively. After the second measurement, we have to make a distinction according to whether we want to take into account correlations between the subsystems or not.

If we disregard the correlations, the ensemble average over all the local measurement outcomes of the state of the quantum system at $t = \tau^{+}$ is described by the following product state $\rho_{A,\tau}\otimes\rho_{B,\tau}$, where
\begin{equation}
\begin{cases}
\rho_{A,\tau}=\sum_{k}p(a_{k}^{\textrm{fin}})\Pi^{\textrm{fin}}_{A,k} \\
\rho_{B,\tau}=\sum_{l}p(b_{l}^{\textrm{fin}})\Pi^{\textrm{fin}}_{B,l} \\
\end{cases}.
\end{equation}
The probabilities $p(a_{k}^{\textrm{fin}})$ to obtain outcome $a_{k}^{\textrm{fin}}$ and
$p(b_{l}^{\textrm{fin}})$ to obtain the measurement outcome $b_{l}^{\textrm{fin}}$ are given by
\begin{equation}
\begin{cases}
p(a_{k}^{\textrm{fin}}) = \textrm{Tr}_{A}\left[\Pi^{\textrm{fin}}_{A,k}\textrm{Tr}_{B}\left[\rho_{\textrm{fin}}\right]\right] \\
p(b_{l}^{\textrm{fin}}) = \textrm{Tr}_{B}\left[\Pi^{\textrm{fin}}_{B,l}\textrm{Tr}_{A}\left[\rho_{\textrm{fin}}\right]\right] \\
\end{cases},
\end{equation}
where $\textrm{Tr}_{A}\left[\cdot\right]$ and $\textrm{Tr}_{B}\left[\cdot\right]$ denote, respectively, the operation of partial trace with respect to the quantum systems $A$ and $B$. Conversely, in order to keep track of the correlations between the simultaneously performed local measurements, we have to take into account the following global observable of the composite system $A-B$:
\begin{equation}\label{eq:ovservable_A-B}
\mathcal{O}^{\textrm{fin}}_{A - B} = \sum_{k,l}c_{kl}^{\textrm{fin}}\Pi^{\textrm{fin}}_{A - B,kl}~,
\end{equation}
where $\Pi^{\textrm{fin}}_{A - B,kl} \equiv \Pi^{\textrm{fin}}_{A,k}\otimes \Pi^{\textrm{fin}}_{B,l}$ and $\{c^{\textrm{fin}}_{kl}\}$ are the outcomes of the final measurement of the protocol.
The state of the system after the second measurement at $t = \tau^{+}$ is then described by an ensemble average over all outcomes of the joint measurements:
\begin{equation}
\rho_{\tau} = \sum_{k,l}p(c^{\textrm{fin}}_{kl})\Pi^{\textrm{fin}}_{A - B,kl}~,
\end{equation}
where $p(c^{\textrm{fin}}_{kl}) = \textrm{Tr}\left[\Pi^{\textrm{fin}}_{A - B,kl}~\rho_{\textrm{fin}}\right]$. In both cases, consistently with the previous sections, we choose $\rho_{\tau}$ as the reference state of the composite system. The measurement outcomes of the initial and final measurement for the composite system $A - B$ are, respectively, $c^{\textrm{in}}_{mh}\equiv(a_{m}^{\textrm{in}},b_{h}^{\textrm{in}})$ and $c^{\textrm{fin}}_{kl}\equiv(a_{k}^{\textrm{fin}},b_{l}^{\textrm{fin}})$. These outcomes occur with probabilities $p(c^{\textrm{in}}_{mh})$ and $p(c^{\textrm{fin}}_{kl})$, which reflect the correlation of the outcomes of the local measurements. As a result, the stochastic quantum entropy production of the composite system reads
\begin{equation}
\sigma_{A - B}(c^{\textrm{in}}_{mh},c^{\textrm{fin}}_{kl}) = \ln\left[\frac{p(c^{\textrm{in}}_{mh})}{p(c^{\textrm{fin}}_{kl})}\right],
\end{equation}
consistently with the definition in section 2. Under the same hypotheses, we can define similar contributions of the stochastic quantum entropy production separately for each subsystem, {\em i.e.} $\sigma_A$ for subsystem $A$ and $\sigma_B$ for subsystem $B$:
\begin{equation}
 \sigma_A(a^{\textrm{in}}_{m},a^{\textrm{fin}}_{k})=\ln\left[\frac{p(a^{\textrm{in}}_{m})}{p(a^{\textrm{fin}}_{k})}\right],~~~~\text{and}~~~~
 \sigma_B(b^{\textrm{in}}_{h},b^{\textrm{fin}}_{l})=\ln\left[\frac{p(b^{\textrm{in}}_{h})}{p(b^{\textrm{fin}}_{l})}\right].
\end{equation}
If upon measurement the composite system is in a product state, the measurement outcomes for $A$ and $B$ are independent and the probabilities to obtain them factorize as
\begin{equation*}
\begin{cases}
p(c^{\textrm{in}}_{mh}) = p(a^{\textrm{in}}_{m})p(b^{\textrm{in}}_{h}) \\
p(c^{\textrm{fin}}_{kl}) = p(a^{\textrm{fin}}_{k})p(b^{\textrm{fin}}_{l})
\end{cases}.
\end{equation*}
As a direct consequence, the stochastic quantum entropy production becomes an additive quantity:
\begin{equation}\label{eq:sigma_AB}
\sigma_{A - B}(c^{\textrm{in}}_{mh},c^{\textrm{fin}}_{kl}) = \sigma_A(a^{\textrm{in}}_{m},a^{\textrm{fin}}_{k}) + \sigma_B(b^{\textrm{in}}_{h},b^{\textrm{fin}}_{l}) \equiv \sigma_{A + B}(c^{\textrm{in}}_{mh},c^{\textrm{fin}}_{kl}).
\end{equation}
In the more general case of correlated measurement outcomes, instead, the probabilities do not factorize anymore, and Eq.~(\ref{eq:sigma_AB}) is not valid anymore. In particular, the mean value of the stochastic entropy production $\sigma_{A - B}(c^{\textrm{in}}_{mh},c^{\textrm{fin}}_{kl})$ becomes sub-additive. In other words
\begin{equation}\label{eq:sigma_sum}
\langle\sigma_{A - B}\rangle \leq \langle\sigma_{A}\rangle + \langle\sigma_{B}\rangle \equiv \langle\sigma_{A + B}\rangle,
\end{equation}
{\em i.e.} the mean value of the stochastic quantum entropy production $\sigma_{A - B}$ of the composite system $A - B$ is smaller than the sum of the mean values of the corresponding entropy production of its subsystems, when the latter are correlated. To see this, we recall the expression of the mean value of the stochastic entropy production in terms of the von Neumann entropies of the two post-measurement states (see \ref{appendix_b}):
\begin{eqnarray*}
\langle\sigma_{A - B}\rangle &=& S(\rho_{\tau})-S(\rho_{\textrm{in}})
=S(\rho_{\tau})-S(\rho_{A,\textrm{in}})-S(\rho_{B,\textrm{in}})\\
&\leq& S(\rho_{A,\tau})+S(\rho_{B,\tau})-S(\rho_{A,\textrm{in}})-S(\rho_{B,\textrm{in}})\\
&=& \langle\sigma_{A}\rangle + \langle\sigma_{B}\rangle = \langle\sigma_{A + B}\rangle.
\end{eqnarray*}

In the following we will analyze the probability distribution of the stochastic quantum entropy productions $\sigma_{A-B}$ for the composite system and
$\sigma_A$, $\sigma_B$ for the subsystems. For comparison we compute also
$\sigma_{A+B}$. We will show in particular how to calculate the corresponding characteristic functions. In the next section we will then show how these characteristic functions can be measured and how they can be used to reconstruct the probability distributions.

\subsection{Probability Distribution}

Depending on the values assumed by the measurement outcomes $c^{\textrm{in}}\in\{c^{\textrm{in}}_{mh}\}$ and $c^{\textrm{fin}}\in \{c^{\textrm{fin}}_{kl}\}$, $\sigma_{A - B}$ is a fluctuating variable as it is true also for the single subsystem contributions $\sigma_A\in\{\sigma_A(a^{\textrm{in}}_{m},a^{\textrm{fin}}_{k})\}$ and $\sigma_B\in\{\sigma_B(b^{\textrm{in}}_{h},b^{\textrm{fin}}_{l})\}$. We denote the probability distributions for the subsystems with $\textrm{Prob}(\sigma_A)$ and $\textrm{Prob}(\sigma_B)$ and $\textrm{Prob}(\sigma_{A-B})$ for the composite system. We will further compare this probability distribution for the composite system (containing the correlations of the local measurement outcomes) to the uncorrelated distribution of the sum of the single subsystems' contributions. We introduce the probability distribution $\textrm{Prob}(\sigma_{A+B})$ of the stochastic quantum entropy production $\sigma_{A+B}$ by applying the following discrete convolution sum:
\begin{equation}\label{eq:convolution}
\textrm{Prob}(\sigma_{A+B}) = \sum_{\{\xi_{B}\}}\textrm{Prob}((\sigma_{A+B} - \xi_{B})_{A})\textrm{Prob}(\xi_{B}),
\end{equation}
where $(\sigma_{A+B} - \xi_{B})_{A}$ and $\xi_{B}$ belong, respectively, to the sample space ({\em i.e.} the set of all possible outcomes) of the random variables $\sigma_A$ and $\sigma_B$.

The probability distribution for the single subsystem, {\em e.g.} the subsystem $A$, is fully determined by the knowledge of the measurement outcomes and the respective probabilities (relative frequencies). We obtain the measurement outcomes $(a^{\textrm{in}}_m,a^{\textrm{fin}}_k)$ with a certain probability $p_{a}(k,m)$, the joint probability for $a^{\textrm{in}}_m$ and $a^{\textrm{fin}}_k$, and this measurement outcome yields the stochastic entropy production $\sigma_A = \sigma_A(a^{\textrm{in}}_{m},a^{\textrm{fin}}_{k})$. Likewise, for system $B$ we introduce the joint probability $p_{b}(l,h)$ to obtain $(b^{\textrm{in}}_h,b^{\textrm{fin}}_l)$, which yields $\sigma_B = \sigma_B(b^{\textrm{in}}_{h},b^{\textrm{fin}}_{l})$.
Therefore, the probability distributions $\textrm{Prob}(\sigma_A)$ and $\textrm{Prob}(\sigma_B)$ are given by
\begin{equation}\label{prob_a}
\textrm{Prob}(\sigma_A) = \left\langle\delta\left[\sigma_A - \sigma_A(a^{\textrm{in}}_{m},a^{\textrm{fin}}_{k})\right]\right\rangle
= \sum_{k,m}\delta\left[\sigma_A - \sigma_A(a^{\textrm{in}}_{m},a^{\textrm{fin}}_{k})\right]p_{a}(k,m)
\end{equation}
and
\begin{equation}\label{prob_b}
\textrm{Prob}(\sigma_B) = \left\langle\delta\left[\sigma_B - \sigma_B(b^{\textrm{in}}_{h},b^{\textrm{fin}}_{l})\right]\right\rangle
= \sum_{l,h}\delta\left[\sigma_B - \sigma_B(b^{\textrm{in}}_{h},b^{\textrm{fin}}_{l})\right]p_{b}(l,h),
\end{equation}
where $\delta[\cdot]$ is the Dirac-delta distribution. In Eqs.~(\ref{prob_a}) and (\ref{prob_b}), the joint probabilities $p_{a}(k,m)$ and $p_{b}(l,h)$ read
\begin{equation}\label{joint}
\begin{cases}
p_{a}(k,m) = \textrm{Tr}\left[(\Pi^{\textrm{fin}}_{A,k}\otimes\mathbbm{1}_{B})\Phi(\Pi^{\textrm{in}}_{A,m}\otimes\rho_{\textrm{B,in}})\right]p(a_{m}^{\textrm{in}}) \\
p_{b}(l,h) = \textrm{Tr}\left[(\mathbbm{1}_{A}\otimes\Pi^{\textrm{fin}}_{B,l})\Phi(\rho_{\textrm{A,in}}\otimes\Pi^{\textrm{in}}_{B,h})\right]p(b_{h}^{\textrm{in}}).
\end{cases}
\end{equation}

By definition, given the reconstructed probability distributions $\textrm{Prob}(\sigma_A)$ and $\textrm{Prob}(\sigma_B)$, the probability $\textrm{Prob}(\sigma_{A+B})$ can be calculated straightforwardly by calculating the convolution of $\textrm{Prob}(\sigma_A)$ and $\textrm{Prob}(\sigma_B)$ according to Eq.~(\ref{eq:convolution}).
Equivalently, the probability distribution $\textrm{Prob}(\sigma_{A - B})$ of the stochastic quantum entropy production of the composite system (containing the correlations between the local measurement outcomes) is given by:
\begin{equation}
\textrm{Prob}(\sigma_{A - B}) = \left\langle\delta\left[\sigma_{A - B} - \sigma_{A - B}(c^{\textrm{in}}_{mh},c^{\textrm{fin}}_{kl})\right]\right\rangle
=\sum_{mh,kl}\delta\left[\sigma_{A - B} - \sigma_{A - B}(c^{\textrm{in}}_{mh},c^{\textrm{fin}}_{kl})\right]p_{c}(mh,kl),
\end{equation}
where
\begin{equation}
p_{c}(mh,kl) = \textrm{Tr}\left[\Pi^{\textrm{fin}}_{A - B,kl}\Phi\left(\Pi^{\textrm{in}}_{A,m}\otimes\Pi^{\textrm{in}}_{B,h}\right)\right]p(c^{\textrm{in}}_{mh}),
\end{equation}
with $p(c^{\textrm{in}}_{mh}) = p(a^{\textrm{in}}_{m})p(b^{\textrm{in}}_{h})$. Now, the integral quantum fluctuation theorems for $\sigma_A$, $\sigma_B$ and $\sigma_{A - B}$ can be derived just by computing the characteristic functions of the corresponding probability distributions $\textrm{Prob}(\sigma_A)$, $\textrm{Prob}(\sigma_B)$ and $\textrm{Prob}(\sigma_{A - B})$, as it will be shown below.

\subsection{Characteristic function of the stochastic quantum entropy production and integral fluctuation theorem}

In probability theory, the characteristic function of a real-valued random variable is its Fourier transform and completely defines the properties of the corresponding probability distribution in the frequency domain~\cite{Kallenberg2005}. We define the characteristic function $G_C(\lambda)$ of the probability distribution $\textrm{Prob}(\sigma_C)$ (for $C\in \{A,B,A-B\}$) as
\begin{equation}
G_C(\lambda) = \int \textrm{Prob}(\sigma_C)e^{i\lambda\sigma_C} d\sigma_C,
\end{equation}
where $\lambda\in\mathbb{C}$ is a complex number. For the two subsystems, by inserting Eqs.~(\ref{prob_a})-(\ref{joint}) and exploiting the linearity of the CPTP quantum maps and of the trace, as shown in \ref{appendix_charact_func}, the characteristic functions for $\textrm{Prob}(\sigma_A)$ and $\textrm{Prob}(\sigma_B)$ can be written as
\begin{small}
\begin{equation}\label{eq:G_A}
G_A(\lambda) = \textrm{Tr}\left\{\left[(\rho_{A,\tau})^{-i\lambda}\otimes\mathbbm{1}_{B}\right]\Phi\left[
(\rho_{\textrm{A,in}})^{1 + i\lambda}\otimes\rho_{\textrm{B,in}}\right]\right\}
\end{equation}
\end{small}
and
\begin{small}
\begin{equation}\label{eq2}
G_B(\lambda) = \textrm{Tr}\left\{\left[\mathbbm{1}_{A}\otimes(\rho_{B,\tau})^{-i\lambda}\right]\Phi\left[
\rho_{\textrm{A,in}}\otimes(\rho_{B,\textrm{in}})^{1 + i\lambda}\right]\right\}.
\end{equation}
\end{small}
In a similar way, we can derive the characteristic function $G_{A - B}(\lambda)$ of the stochastic entropy production of the composite system $A - B$:
\begin{equation}\label{eq:G_A-B}
G_{A - B}(\lambda) = \mathrm{Tr}\left[\rho_{\tau}^{-i\lambda}\Phi(\rho_{\mathrm{in}}^{1+i\lambda})\right]\,.
\end{equation}
Furthermore, if we choose $\lambda = i$, the integral quantum fluctuation theorems can be straightforwardly derived, namely for $\sigma_A$ and $\sigma_B$:
\begin{small}
\begin{equation}\label{eq3}
\big\langle e^{-\sigma_A}\big\rangle \equiv G_A(i) = \textrm{Tr}\left\{\left[\rho_{A,\tau}\otimes\mathbbm{1}_{B}\right]\Phi\left[
\mathbbm{1}_{A}\otimes\rho_{\textrm{B,in}}\right]\right\}
\end{equation}
\end{small}%
and
\begin{small}
\begin{equation}\label{eq4_bis}
\big\langle e^{-\sigma_B}\big\rangle \equiv G_B(i) = \textrm{Tr}\left\{\left[\mathbbm{1}_{A}\otimes\rho_{B,\tau}\right]\Phi\left[
\rho_{\textrm{A,in}}\otimes\mathbbm{1}_{B}\right]\right\},
\end{equation}
\end{small}%
as well as
\begin{small}
\begin{equation}\label{eq4}
\big\langle e^{-\sigma_{A-B}}\big\rangle \equiv G_{A-B}(i) = \textrm{Tr}\left\{\rho_{\tau}\Phi\left[
\mathbbm{1}_{A-B}\right]\right\}=1
\end{equation}
\end{small}%
for $\sigma_{A-B}$ (with $\Phi$ unital). The characteristic functions of Eqs.~(\ref{eq:G_A})-(\ref{eq:G_A-B}) depend exclusively on appropriate powers of the initial and final density operators of each subsystem. These density operators are diagonal in the basis of the observable eigenvectors and can be measured by means of standard state population measurements for each value of $\lambda$. As will be shown in the following, this result can lead to a significant reduction of the number of measurements that is required to reconstruct the probability distribution of the stochastic quantum entropy production, beyond the direct application of the definition of Eqs.~(\ref{prob_a})-(\ref{joint}). A reconstruction algorithm implementing such improvement will be discussed in the next section.

\section{Reconstruction algorithm}
\label{VVV}

In this section, we present the algorithm for the reconstruction of the probability distribution $\textrm{Prob}(\sigma)$ for the stochastic quantum entropy production $\sigma$. The procedure is based on a parametric version of the integral quantum fluctuation theorem, {\em i.e.} $\langle e^{-\varphi \sigma}\rangle$ ($\varphi\in\mathbb{R}$). In particular, we introduce the moment generating functions $\chi_{C}(\varphi)$ for $C\in\{A,B,A-B\}$:
$$
\langle e^{-\varphi\sigma_C}\rangle = G_C(i\varphi) \equiv \chi_{C}(\varphi).
$$
The quantity $\langle e^{-\varphi\sigma_{C}}\rangle$ can be expanded into a Taylor series, so that
\begin{equation}
\chi_{C}(\varphi) = \langle e^{-\varphi\sigma_{C}}\rangle = \left\langle\sum_{k}\frac{(-\varphi^{k})}{k!}\sigma_{C}^{k}\right\rangle
= 1 - \varphi\langle\sigma_{C}\rangle + \frac{\varphi^{2}}{2}\langle\sigma_{C}^{2}\rangle - \ldots
\end{equation}
Accordingly, the statistical moments of the stochastic quantum entropy production $\sigma_{C}$, denoted by $\{\langle\sigma_{C}^{k}\rangle\}$ with $k = 1,\ldots,N-1$, can be expressed in terms of the $\chi_{C}(\varphi)$'s defined over the parameter vector $\underline{\varphi} \equiv (\varphi_{1}, \ldots, \varphi_{N})^T$, {\em i.e.}
\begin{equation}
\begin{pmatrix} \chi_{C}(\varphi_{1}) \\ \chi_{C}(\varphi_{2}) \\ \vdots \\ \chi_{C}(\varphi_{N}) \end{pmatrix} = \underbrace{\begin{pmatrix} 1  & -\varphi_{1} & +\frac{\varphi_{1}^{2}}{2} & \ldots & \frac{\varphi_{1}^{N-1}}{N-1!} \\
1  & -\varphi_{2} & +\frac{\varphi_{2}^{2}}{2} & \ldots & \frac{\varphi_{2}^{N-1}}{N-1!} \\
\vdots & \vdots & \vdots & \vdots & \vdots \\
1  & -\varphi_{N} & +\frac{\varphi_{N}^{2}}{2} & \ldots & \frac{\varphi_{N}^{N-1}}{N-1!} \end{pmatrix}}_{A(\underline{\varphi})}
\begin{pmatrix} 1 \\ \langle\sigma_{C}\rangle \\ \langle\sigma_{C}^{2}\rangle \\ \vdots \\ \langle\sigma_{C}^{N-1}\rangle \end{pmatrix},
\end{equation}
where the matrix $A(\underline{\varphi})$ can be written as a Vandermonde matrix, as detailed below. It is clear at this point that the solution to the problem to infer the set $\{\langle\sigma_{C}^{k}\rangle\}$ can be related to the resolution of a polynomial interpolation problem, where the experimental data-set is given by $N$ evaluations of the parametric integral fluctuation theorem of $\sigma_{C}$ in terms of the $\varphi$'s. Let us observe that only by choosing real values for the parameters $\varphi$ is it possible to set up the proposed reconstruction procedure via the resolution of an interpolation problem. We will explain in the next section a feasible experiment with trapped ions to directly measure the quantities $\chi_{C}(\varphi)$ by properly varying the parameter $\varphi$. By construction, the dimension of the parameters vector $\underline{\varphi}$ is equal to the number of statistical moments of $\sigma_{C}$ that we want to infer, including the trivial zero-order moment. In this regard, we define the vectors
\begin{equation*}
\widetilde{\underline{m}}\equiv\left(1, \, \,
-\langle\sigma_{C}\rangle, \,\,\, \ldots, \,\, (-1)^{N-1}\frac{\langle\sigma_{C}^{N-1}\rangle}{N-1!}\right)^T,
\end{equation*}
with element ${\widetilde{\underline{m}}}_j=(-1)^{j}\frac{\langle\sigma_{C}^{j}\rangle}{j!}$, $j=0, \ldots, N-1$, and
$$
\underline{\chi}_{C}\equiv(\chi_{C}(\varphi_{1}), \ldots, \chi_{C}(\varphi_{N}))^T.
$$
Then one has
\begin{equation}\label{vandermonde}
\underline{\chi}_{C} = V(\underline{\varphi})\widetilde{\underline{m}},
\end{equation}
where
\begin{equation}
V(\underline{\varphi}) = \begin{pmatrix} 1  & \varphi_{1} & \varphi_{1}^{2} & \ldots & \varphi_{1}^{N-1} \\
1  & \varphi_{2} & \varphi_{2}^{2} & \ldots & \varphi_{2}^{N-1} \\
\vdots & \vdots & \vdots & \vdots & \vdots \\
1  & \varphi_{N} & \varphi_{N}^{2} & \ldots & \varphi_{N}^{N-1} \end{pmatrix}
\end{equation}
is the Vandermonde matrix built on the parameters vector $\underline{\varphi}$. $V(\underline{\varphi})$ is a matrix whose rows (or columns) have elements in geometric progression, {\em i.e.} $v_{ij} = \varphi^{j-1}_{i}$, where $v_{ij}$ denotes the $ij-$ element of $V(\underline{\varphi})$. Eq.~(\ref{vandermonde}) constitutes the formula for the inference of the statistical moments $\{\langle\sigma_{C}^{k}\rangle\}$ by means of a finite number $N$ of evaluations of $\chi_{C}(\varphi)$. Moreover, given the vector
$\underline{m}\equiv(1, \,\, \langle\sigma_{C}\rangle,\ldots,\langle\sigma_{C}^{N-1}\rangle)^T$ of the statistical moments of $\sigma_{C}$, the linear transformation ${\cal T}$, which relates $\widetilde{\underline{m}}$ with $\underline{m}$ such that $\underline{m} = {\cal T} \widetilde{\underline{m}}$, is
${\cal T} = \textrm{diag}\left(\{(-1)^{n}n!\}_{n=0}^{N-1}\right)$, where $\textrm{diag}(\cdot)$ denotes the diagonal matrix. The determinant of the Vandermonde matrix $V(\underline{\varphi})$ is
$$
\textrm{det}\left[V(\underline{\varphi})\right] = \prod_{1\leq i\leq j\leq N}(\varphi_{j} - \varphi_{i}),
$$
given by the product of the differences between all the elements of the vector $\underline{\varphi}$, which are counted only once with their appropriate sign. As a result, $\textrm{det}\left[V(\underline{\varphi})\right] = 0$ if and only if $\underline{\varphi}$ has at least two identical elements. Only in that case, the inverse of $V(\underline{\varphi})$ does not exist and the polynomial interpolation problem cannot be longer solved. However, although the solution of a polynomial interpolation by means of the inversion of the Vandermonde matrix exists and is unique, $V(\underline{\varphi})$ is an ill-conditioned matrix~\cite{Meyer2000}. This means that the matrix is highly sensitive to small variations of the set of the input data (in our case the parameters $\varphi$'s), such that the condition number of the matrix may be large and the matrix becomes singular. As a consequence, the reconstruction procedure will be computationally inefficient, especially in the case the measurements are affected by environmental noise. Numerically stable solutions of a polynomial interpolation problem usually rely on the Newton polynomials~\cite{Trefethen2000}. The latter allow us to write the characteristic function $\chi_{C}(\varphi)$ in polynomial terms as a function of each value of $\underline{\varphi}$:
\begin{equation}
\chi^{\textrm{pol}}_{C}(\varphi) = \sum^{N}_{k=1}\eta_{k}n_{k}(\varphi),
\end{equation}
with $n_{k}(\varphi)\equiv\prod^{k-1}_{j=1}(\varphi - \varphi_{j})$ and $n_{1}(\varphi) = 1$. The coefficients $\eta_{k}$ of the basis polynomials, instead, are given by the divided differences
\begin{equation}
\eta_{k} = \left[\chi_{C}(\varphi_{1}),\ldots,\chi_{C}(\varphi_{k})\right]
\equiv \frac{\left[\chi_{C}(\varphi_{2}),\ldots,\chi_{C}(\varphi_{k})\right] -
\left[\chi_{C}(\varphi_{1}),\ldots,\chi_{C}(\varphi_{k-1})\right]}{(\varphi_{k} - \varphi_{1})},
\end{equation}
where $\left[\chi_{C}(\varphi_{k})\right]\equiv\chi_{C}(\varphi_{k})$, $\left[\chi_{C}(\varphi_{k-1}),\chi_{C}(\varphi_{k})\right] \equiv ([\chi_{C}(\varphi_{k})] - [\chi_{C}(\varphi_{k-1})])/(\varphi_{k} - \varphi_{k-1}) = (\chi_{C}(\varphi_{k}) - \chi_{C}(\varphi_{k-1}))/(\varphi_{k} - \varphi_{k-1})$, and all the other divided differences found recursively.

Then, the natural question arises on what is an optimal choice for $\underline{\varphi}$. It is essential, indeed, to efficiently reconstruct the set $\{\langle\sigma_{C}^{k}\rangle\}$ of the statistical moments of $\sigma_{C}$. For this purpose, we can take into account the error $e_{C}(\varphi) \equiv \chi_{C}(\varphi) - \chi^{\textrm{pol}}_{C}(\varphi)$ in solving the polynomial interpolation problem in correspondence of a value of $\varphi$ different from the interpolating points within the parameter vector $\underline{\varphi}$. The error
$e_{C}(\varphi)$ depends on the regularity of the function $\chi_{C}(\varphi)$, and especially on the values assumed by the parameters $\varphi$. As shown in~\cite{Trefethen2000,Phillips2003}, the choice of the $\varphi$'s for which the interpolation error is minimized is given by the real zeros of the Chebyshev polynomial of degree $N$ in the interval $\left[\varphi_{\textrm{min}},\varphi_{\textrm{max}}\right]$, where $\varphi_{\textrm{min}}$ and $\varphi_{\textrm{max}}$ are, respectively, the lower and upper bound of the parameters $\varphi$. Accordingly, the optimal choice for $\underline{\varphi}$ is given by
\begin{equation}\label{cheby}
\varphi_{k} = \frac{(\varphi_{\textrm{min}}+\varphi_{\textrm{max}})}{2} + \frac{\varphi_{\textrm{max}}-\varphi_{\textrm{min}}}{2}\cos\left(\frac{2k-1}{2N}\pi\right),
\end{equation}
with $k = 1,\ldots,N$. Let us observe that the value of $N$, {\em i.e.} the number of evaluations of the characteristic function $\chi_{C}(\varphi)$, is equal to the number of statistical moments of $\sigma_{C}$ we want to infer. Therefore, in principle, if the probability distribution of the stochastic quantum entropy production is a Gaussian function, then $N$ could be taken equal to $2$.

Hence, once all the evaluations of the characteristic functions $\chi_{C}(\varphi)$ have been collected, we can derive the statistical moments of the quantum entropy production $\sigma_{C}$, and consequently reconstruct the probability distribution $\textrm{Prob}(\sigma_{C})$ as
\begin{equation}\label{Fourier}
\textrm{Prob}(\sigma_{C}) \approx \mathcal{F}^{-1}\left\{\sum_{k = 0}^{N-1}\frac{\langle\sigma_{C}^{k}\rangle}{k!}(i\mu)^{k}\right\}
\equiv\frac{1}{2\pi}\int^{\infty}_{-\infty}\left[\sum_{k = 0}^{N-1}\frac{\langle\sigma_{C}^{k}\rangle}{k!}(i\mu)^{k}\right]e^{-i\mu\sigma_{C}}d\mu,
\end{equation}
where $\mu\in\mathbb{R}$ and $\mathcal{F}^{-1}$ denotes the inverse Fourier transform~\cite{Mnatsakanov}, which is numerically performed~\cite{Athanassoulis}. To do that, we fix a-priori the integration step $d\mu$ and we vary the integration limits of the integral, in order to minimize the error
$\sum_{k}\left|\widetilde{\langle\sigma_{C}^{k}\rangle} - \overline{\langle\sigma_{C}^{k}\rangle}\right|^{2}$ between the statistical moments $\widetilde{\langle\sigma_{C}^{k}\rangle}$, obtained by measuring the characteristic functions $\chi_{C}(\varphi)$ (i.e. after the inversion of the Vandermonde matrix), and the ones calculated from the reconstructed probability distribution, $\overline{\langle\sigma_{C}^{k}\rangle}$, which we derive by numerically computing the inverse Fourier transform for each value of $\sigma_{C}$. This procedure has to be done separately for $C\in\{A,B,A-B\}$, while, as mentioned, the probability distribution $\textrm{Prob}(\sigma_{A+B})$ is obtained by a convolution of $\textrm{Prob}(\sigma_{A})$ and $\textrm{Prob}(\sigma_{B})$. Here, it is worth observing that Eq. (\ref{Fourier}) provides an approximate expression for the probability distribution $\textrm{Prob}(\sigma_{C})$. Ideally, given a generic unital quantum CPTP map modeling the dynamics of the system, an infinite number $N$ of statistical moment of $\sigma_{C}$ is required to reconstruct $\textrm{Prob}(\sigma_{C})$ if we use the inverse Fourier transform as in Eq. (\ref{Fourier}).
\begin{figure*}[t]
	\centering
	\includegraphics[scale=3.25]{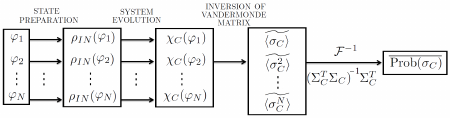}
	\caption{\textbf{Pictorial representation of the reconstruction algorithm -} The reconstruction algorithm starts by optimally choosing the parameters $\varphi\in\{\alpha,\beta,\gamma\}$ as the zeros of the Chebyshev polynomial of degree $N$ in the intervals $[\varphi_{\textrm{min}},\varphi_{\textrm{max}}]$. Then, the moment generating functions $\chi_{C}(\varphi)$, with $C\in\{A,B,A-B\}$, are measured. The experimental steps for their measuring and a detail analysis about the required number of measurements to perform the procedure are explained in the main text. Once the estimates {\footnotesize $\widetilde{\langle \sigma_C^k\rangle}$} of the statistical moments of $\sigma_{C}$ are obtained, the inverse Fourier transform $\mathcal{F}^{-1}$ has to be numerically performed. Alternatively, the Moore-Penrose pseudo-inverse of $\Sigma_{C}$ can be adopted. As a result, an estimate {\footnotesize $\overline{\textrm{Prob}(\sigma_{C})}$} for the probability distribution $\textrm{Prob}(\sigma_{C})$ is obtained.}
	\label{fig:procedure}
\end{figure*}
While we can always calculate the Fourier transform to reconstruct the probability distribution from its moments, in the case of a distribution with discrete support (as in our case), there is a different method that can lead to higher precision, especially when the moment generating function is not approximated very well by the chosen number $N$ of extracted moments. As a matter of fact, each statistical moment $\widetilde{\langle\sigma_{C}^{k}\rangle}$, with $C\in\{A,B,A-B\}$, is the best approximation of the true statistical moments of $\sigma_{C}$ from the measurement of the corresponding characteristic functions $\chi_C(\varphi)$. Hence, apart from a numerical error coming from the inversion of the Vandermonde matrix $A$ or the use of the Newton polynomials $\chi^{\textrm{pol}}_{C}$, we can state that
\begin{equation}\label{expansion_sigma_C}
\widetilde{\langle\sigma_{C}^{k}\rangle} \simeq \sum_{i = 1}^{M_{C}}\sigma^{k}_{C,i}\textrm{Prob}(\sigma_{C,i})
= \sigma_{C,1}^{k}\textrm{Prob}(\sigma_{C,1}) + \ldots + \sigma_{C,M_{C}}^{k}\textrm{Prob}(\sigma_{C,M_{C}}),
\end{equation}
with $k = 1,\ldots,N$. In Eq.~(\ref{expansion_sigma_C}), $M_{C}$ is equal to the number of values that can be assumed by $\sigma_{C}$, while $\sigma_{C,i}$ denotes the $i-$th possible value for the stochastic quantum entropy production of the (sub)system $C$. As a result, the probabilities $\textrm{Prob}(\sigma_{C,i})$, $i = 1,\ldots,M$, can be approximately expressed as a function of the statistical moments $\left\{\widetilde{\langle\sigma_{C}^{k}\rangle}\right\}$, {\em i.e.}
\begin{equation}
\begin{pmatrix} \widetilde{\langle\sigma_{C}\rangle} \\ \widetilde{\langle\sigma_{C}^{2}\rangle} \\ \vdots \\ \widetilde{\langle\sigma_{C}^{N}\rangle} \end{pmatrix} = \underbrace{\begin{pmatrix} \sigma_{C,1}  & \sigma_{C,2} & \ldots & \sigma_{C,M} \\
\sigma_{C,1}^{2}  & \sigma_{C,2}^{2} & \ldots & \sigma_{C,M}^{2} \\
\vdots & \vdots & \vdots & \vdots \\
\sigma_{C,1}^{M}  & \sigma_{C,2}^{M} &  \ldots & \sigma_{C,N}^{M} \end{pmatrix}}_{\Sigma_{C}}
\begin{pmatrix} \textrm{Prob}(\sigma_{C,1}) \\ \textrm{Prob}(\sigma_{C,2}) \\ \vdots \\ \textrm{Prob}(\sigma_{C,M}) \end{pmatrix},
\end{equation}
where $\Sigma_{C}\in\mathbb{R}^{N\times M}$. By construction $\Sigma_{C}$ is a rectangular matrix, that is computed by starting from the knowledge of the values assumed by the stochastic quantum entropy production $\sigma_{C,i}$. Finally, in order to obtain the probabilities $\textrm{Prob}(\sigma_{C,i})$, $i = 1,\ldots,M_C$, we have to adopt the Moore-Penrose pseudo-inverse of $\Sigma_{C}$, which is defined as
\begin{equation}
\Sigma^{+}_{C}\equiv (\Sigma^{T}_{C}\Sigma_{C})^{-1}\Sigma^{T}_{C}.
\end{equation}

A pictorial representation of the reconstruction protocol is shown in Fig.~\ref{fig:procedure}. Let us observe, again, that the proposed algorithm is based on the expression of Eq.~(\ref{sigma}) for the stochastic quantum entropy production, which has been obtained by assuming unital CPTP quantum maps for the system dynamics. We expect that for a general open quantum system, not necessarily described by a unital CPTP map, one can extend the proposed reconstruction protocol, even though possibly at the price of a greater number of measurements. Notice that, since Eq.~(\ref{equality_cond_prob}) it is not longer valid in the general case, one has to use directly Eqs.~(\ref{general_sigma})-(\ref{general_mean_sigma}). However, we observe that, as shown in \cite{Manzano2015}, the ratio between the conditional probabilities
may admit for a large family of CPTP maps the form $p(a^{\textrm{fin}}_{k}|a^{\textrm{in}}_{m}) / p(a^{\textrm{in}}_{m}|a^{\textrm{ref}}_{k})
\equiv e^{-\Delta V}$, where the quantity $\Delta V$ is related to the so-called nonequilibrium potential, so that
$\sigma=\sigma_{unital}+V$ and $\sigma_{unital}$ again given by Eq.~(\ref{sigma}).

\subsection{Required number of measurements}

From an operational point of view, we need to measure (directly or indirectly) the quantities
\begin{equation}
\begin{cases}
\chi_{A}(\alpha) = \textrm{Tr}\left\{\left[(\rho_{A,\tau})^{\alpha}\otimes\mathbbm{1}_{B}\right]\Phi\left[
(\rho_{\textrm{A,in}})^{1 - \alpha}\otimes\rho_{\textrm{B,in}}\right]\right\} \\
\chi_{B}(\beta) = \textrm{Tr}\left\{\left[\mathbbm{1}_{A}\otimes(\rho_{B,\tau})^{\beta}\right]\Phi\left[
\rho_{\textrm{A,in}}\otimes(\rho_{\textrm{B,in}})^{1 - \beta}\right]\right\}\\
\chi_{A-B}(\gamma) = \textrm{Tr}\left\{(\rho_{\tau})^{\gamma}\Phi\left[
(\rho_{\textrm{in}})^{1 - \gamma}\right]\right\}
\end{cases},
\end{equation}
{\em i.e.} the moment generating functions of $\sigma_A$, $\sigma_B$ and $\sigma_{A-B}$, after a proper choice of the parameters $\alpha$, $\beta$ and $\gamma$, with $\alpha,\beta, \gamma \in\mathbb{R}$. The optimal choice for these parameters was analyzed in the previous section. For this purpose, as shown in \ref{appendix_charact_func}, it is worth mentioning that $\left(\rho_{C,\textrm{in}}\right)^{1-\varphi}\equiv\sum_{m}\Pi^{\textrm{in}}_{C,m}p(x_{m}^{\textrm{in}})^{1 - \varphi}$ and $\left(\rho_{C,\tau}\right)^{\varphi}\equiv\sum_{k}\Pi^{\tau}_{C,k}p(x^{\tau}_{k})^{\varphi}$, where $C\in\{A,B,A - B\}$, $x\in\{a,b,c\}$ and $\varphi\in\{\alpha,\beta,\gamma\}$. A direct measurement of $\chi_{C}(\varphi)$, based for example on an interferometric setting as shown in Ref.~\cite{MazzolaPRL2013,DornerPRL2013} for the work distribution inference, is not trivial, especially for the general fully quantum case. For this reason, we propose a procedure, suitable for experimental implementation, requiring a limited number of measurements, based on the following steps:
\begin{enumerate}
\item
Prepare the initial product state $\rho_{\textrm{in}}=\rho_{A,\textrm{in}}\otimes\rho_{B,\textrm{in}}$, as given in Eq.~(\ref{rho_in}), with fixed probabilities $p(a^{\textrm{in}}_{m})$ and $p(b^{\textrm{in}}_{h})$. Then, after the composite system $A - B$ is evolved within the time interval $[0,\tau]$, measure the occupation probabilities $p(a^{\textrm{fin}}_{k})$ and $p(b^{\textrm{fin}}_{l})$ via local measurements on $A$ and $B$. Then, compute the stochastic quantum entropy productions $\sigma_{A}(a_m^{\textrm{in}},a_k^{\textrm{fin}})$ and $\sigma_{B}(b_h^{\textrm{in}},b_l^{\textrm{fin}})$. Simultaneous measurements on $A$ and $B$ yield also the probabilities $p(c_{kl}^{\textrm{fin}})$ and thus $\sigma_{A-B}(c_{mh}^{\textrm{in}},c_{kl}^{\textrm{fin}})$.
\item
For every chosen value of $\alpha$, $\beta$ and $\gamma$, prepare, for instance by quantum optimal control tools~\cite{DoriaPRL2011}, the quantum subsystems in the states
\begin{equation*}
\begin{cases}
\displaystyle{\rho_{\textrm{IN}}(\alpha)\equiv\frac{\left[(\rho_{A,\textrm{in}})^{1 - \alpha}\otimes\rho_{B,\textrm{in}}\right]}{\textrm{Tr}\left[(\rho_{A,\textrm{in}})^{1 - \alpha}\otimes\rho_{B,\textrm{in}}\right]}} \\
\displaystyle{\rho_{\textrm{IN}}(\beta)\equiv\frac{\left[\rho_{\textrm{A,in}}\otimes(\rho_{B,\textrm{in}})^{1 - \beta}\right]}{\textrm{Tr}\left[\rho_{\textrm{A,in}}\otimes(\rho_{B,\textrm{in}})^{1 - \beta}\right]}}\\
\displaystyle{\rho_{\textrm{IN}}(\gamma)\equiv\frac{\left(\rho_{\textrm{A,in}}\otimes\rho_{B,\textrm{in}}\right)^{1 - \gamma}}{\textrm{Tr}\left[\left(\rho_{\textrm{A,in}}\otimes\rho_{B,\textrm{in}}\right)^{1 - \gamma}\right]}}
\end{cases},
\end{equation*}
and let the system evolve.
\item
Since the characteristic function $\chi_{C}(\varphi)$, with $C\in\{A,B,A-B\}$ and $\varphi\in\{\alpha,\beta,\gamma\}$, is given by performing a trace operation with respect to the composite system $A - B$, one can write the following simplified relation:
\begin{equation}
\chi_{C}(\varphi) = \sum_{k}\sum_{m}\langle m|p(x^{\textrm{fin}}_{k})^{\varphi}|k\rangle\langle k|\rho_{\textrm{FIN}}(\varphi)|m\rangle
= \sum_{m}p(x^{\textrm{fin}}_{m})^{\varphi}\langle m|\rho_{\textrm{FIN}}(\varphi)|m\rangle,
\end{equation}
where $\{|l\rangle\}$, $l = m,k$, is the orthonormal basis of the composite system $A - B$, $x\in\{a,b,c\}$ and $\rho_{\textrm{FIN}}(\varphi)\equiv\Phi[\rho_{\textrm{IN}}(\varphi)]$ (with $p(x^{\textrm{fin}}_{m})$ measured in step 1 and $\rho_{\textrm{IN}}(\varphi)$ introduced in step 2). Thus, measure the occupation probabilities $\langle m|\rho_{\textrm{FIN}}(\varphi)|m\rangle$ in order to obtain all the characteristic functions $\chi_{C}(\varphi)$.
\end{enumerate}
We observe that the measure of the characteristic functions $\chi_{C}(\varphi)$ relies only on the measure of occupation probabilities. Hence, the proposed procedure does not require any tomographic measurement. Moreover, for the three steps of the protocol we can well quantify the required number of measurements to properly infer the statistics of the quantum entropy production regarding the composite quantum system. The required number of measurements, indeed, scales linearly with the number of possible measurement outcomes coming from each quantum subsystem at the initial and final stages of the protocol. Equivalently, if we define $d_{A}$ and $d_{B}$ as the dimension of the Hilbert space concerning the quantum subsystems $A$ and $B$, we can state that the number of measurements for both of the three steps scales linearly with $(d_{A} + d_{B})$, {\em i.e.} with the number of values $(M_{A} + M_{B})$ that can be assumed by $\sigma_{A}$ and $\sigma_{B}$, the stochastic quantum entropy production of the subsystems. It also
scale linearly with $M_A M_B$ for the reconstruction of the stochastic quantum entropy production $\sigma_{A-B}$ of the composite system. The reason is that the described procedure is able to reconstruct the distribution of the stochastic quantum entropy production, without directly measuring the joint probabilities $p_a(k,m)$ and $p_{b}(l,h)$ for the two subsystems and $p_c(mh,kl)$ for the composite system. Otherwise, the number of required measurements would scale, respectively, as $M_{A}^{2}$ and $M_{B}^{2}$ for the subsystems and as $(M_AM_B)^2$ for the composite system in order to realize all the combinatorics concerning the measurement outcomes.

\section{A physical example}
\begin{figure}[t]
	\centering
	\includegraphics[scale=.35]{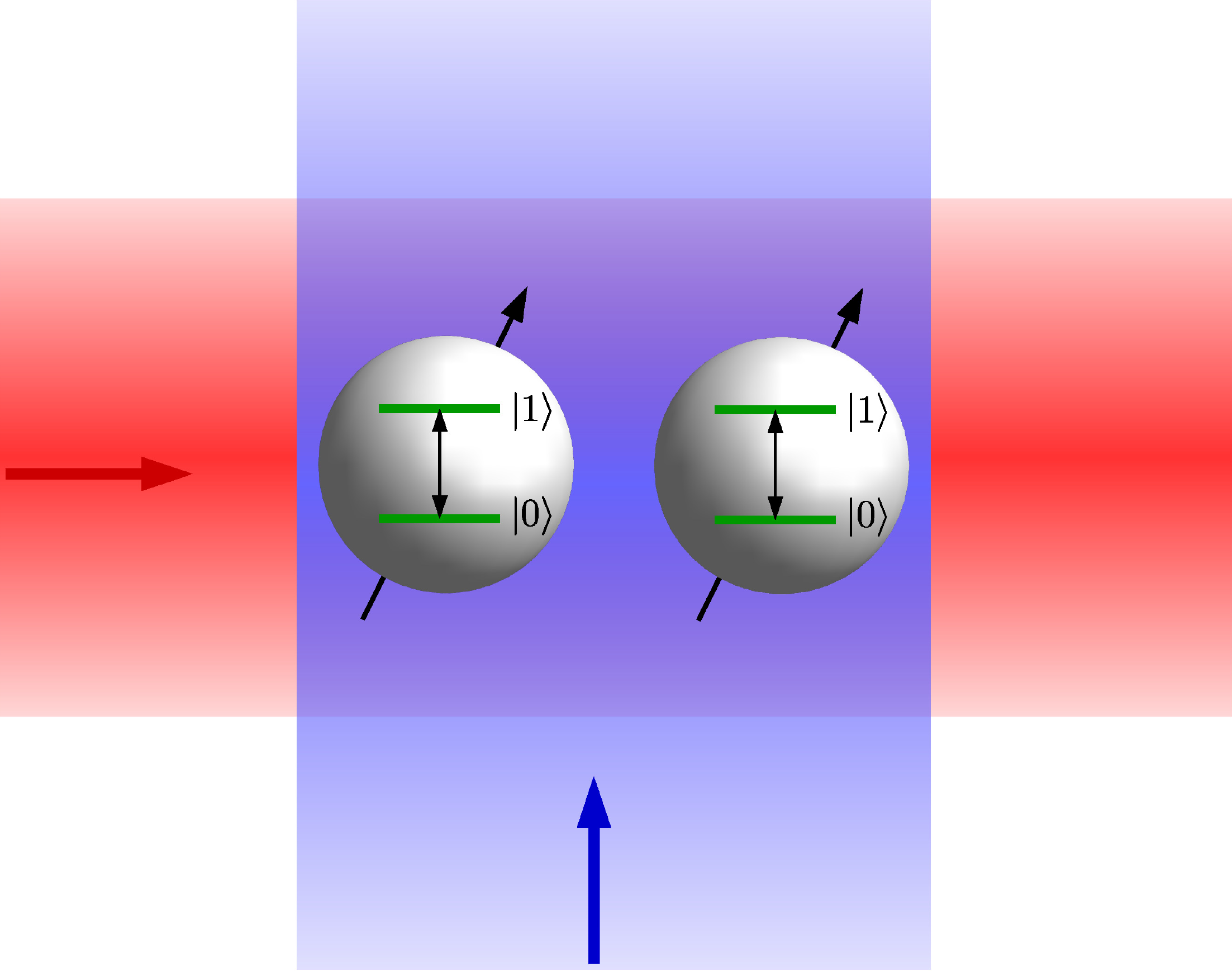}
	\caption{\textbf{Experimental implementation with trapped ions -} Pictorial representation of two trapped ions subjected to two laser fields. The internal levels of the ions allow to encode one qubit in each ion. The transition between these levels is driven by the lasers, where the driving depends on the state of the common vibrational (trap) mode of the two ions. The lasers can be focused to choose between single or global addressing. This allows to generate local gates as well as entangling gates.}
	\label{fig:fig2}
\end{figure}
In the previous section, we have introduced an algorithm for the reconstruction of thermodynamical quantities in a fully quantum regime. Here, in order to illustrate our theoretical results, we discuss in this section an experimental implementation with trapped ions.
Trapped ions have been demonstrated to be a versatile tool for quantum simulation~\cite{Friedenauer2008,Kim2010,Lanyon2011,GerritsmaNAT2010}, including simulation of quantum thermodynamics~\cite{Huber2008,An2015,AbahPRL2012,RossnagelPRL2014,Rossnagel325}. The application of our protocol on a physical example relies on the availability of experimental procedures for state preparation and readout, as well as an entangling operation.

We consider a system of two trapped ions, whose two internal states allow to encode the qubit states $|0\rangle$ and $|1\rangle$ of the standard computational basis. Then, the subsystems $A$ and $B$ are represented by the two qubits. The latter can interact by the common vibrational (trap) mode of the two ions, and external lasers allow to manipulate the ion states, generating arbitrary single qubit rotations through individual addressing or an entangling operation, as for example the M\o{}lmer-S\o{}rensen gate operation~\cite{SoerensenPRL82,RoosNJP2008,MonzPRL106,Nigg302}. Fig.~\ref{fig:fig2} shows a pictorial representation of the system. While usually universal state preparation for single qubits is supposed only for pure states, here we have to prepare mixed states. However, once we have prepared a pure state with the right amount of population in the two levels, we can reach the required mixed state by applying a random $Z$ rotation leading to a complete dephasing of the two levels, where $Z$ is the corresponding Pauli matrix. The two-qubit operation, that generates entanglement between $A$ and $B$, is chosen to be a partial M\o{}lmer-S\o{}rensen gate operation, given by the following unitary operation, depending on the phase $\phi$:
\begin{equation}\label{propagator}
\mathcal{U}(\phi)=e^{-i\phi\left(X^{A}\otimes X^{B}\right)},
\end{equation}
where $X^{A}$ and $X^{B}$ equal, respectively, to the Pauli matrix $X$ for the quantum systems $A$ and $B$, and the reduced Planck's constant $\hbar$ is set to unity. In the following (and unless explicitly stated otherwise), we choose $\phi=\frac{\pi}{7}$, and start from the initial state $\rho_{0} = \textrm{diag}\left(\frac{6}{25},\frac{9}{25},\frac{4}{25},\frac{6}{25}\right)$ since this choice leads to a non-Gaussian probability distribution $\textrm{Prob}(\sigma_{A - B})$ of the stochastic quantum entropy production. For the sake of simplicity, we remove the label $A$ and $B$ from the computational basis $\{|0\rangle,|1\rangle\}$ considered for the two subsystems. Thus, the corresponding projectors are $\Pi_{0}\equiv|0\rangle\langle 0|$ and $\Pi_{1}\equiv|1\rangle\langle 1|$, and each ion is characterized by $4$ different values of the stochastic quantum entropy production $\sigma_{C}$, with $C\in\{A,B\}$. As a consequence, the probability distribution $\textrm{Prob}(\sigma_{A - B})$ of the stochastic quantum entropy production for the composite system $A - B$ is defined over a discrete support given by $l$ samples, with $l \leq M_{A}M_{B} = 16$.

\subsection{Correlated measurement outcomes and correlations witness}

In the general case, the outcomes of the second measurement of the protocol are correlated, as in our example, and the stochastic quantum entropy production of the composite system is sub-additive, {\em i.e.} $\langle\sigma_{A-B}\rangle\leq\langle\sigma_{A}\rangle+\langle\sigma_{B}\rangle$. Hence, by adopting the reconstruction algorithm proposed in Fig.~\ref{fig:procedure} we are able to effectively derive the upper bound of $\langle\sigma_{A - B}\rangle$, which defines the thermodynamic irreversibility for the quantum process. In the simulations of this section, we compare the fluctuation profile that we have derived by performing local measurements on the subsystems $A$ and $B$ with the ones that are obtained via a global measurement on the composite system $A - B$, in order to establish the amount of information which is carried by a set of local measurements. Furthermore, we will discuss the changes of the fluctuation profile of the stochastic quantum entropy production both for unitary and noisy dynamics. The unitary operation describing the dynamics of the quantum system is given by Eq.~(\ref{propagator}), while the noisy dynamics can be described by the differential Lindblad (Markovian) equation $\dot{\rho}(t) = \mathcal{L}(\rho(t))$, defined as
\begin{equation}\label{Lindblad}
\dot{\rho}(t) = -i\left[H,\rho\right] - \sum_{C\in\{A,B\}}\Gamma_{C}\left(\{\rho,L_{C}^{\dagger}L_{C}\} - 2L_{C}\rho L_{C}^{\dagger}\right).
\end{equation}
In Eq.~(\ref{Lindblad}), $\rho(t)$ denotes the density matrix describing the composite quantum system $A - B$, $\{\cdot,\cdot\}$ is the anticommutator, $\Gamma_A$ and $\Gamma_B$ (rad/s) are dephasing rates corresponding to $L_{A}\equiv\Pi_{0}\otimes\mathbbm{1}_{B}$ and $L_{B}\equiv\mathbbm{1}_{A}\otimes\Pi_{0}$, pure-dephasing Lindblad operators, where $\mathbbm{1}_{A}$ and $\mathbbm{1}_{B}$ are the identity operators acting, respectively, on the Hilbert spaces of the ions $A$ and $B$. The Hamiltonian of the composite system $A - B$ in Eq.~(\ref{Lindblad}), instead, is given by $$H = \omega\left(X^{A}\otimes X^{B}\right),$$ where the interaction strength $\omega = \phi/\tau$ (rad/s) with $\tau$ kept fixed and chosen equal to $50$ s (leading to a largely relaxed system dynamics), consistently with the unitary operation (\ref{propagator}).
\begin{figure*}[h!]
	\centering
	\includegraphics[scale=6.15]{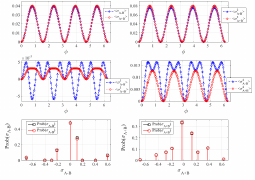}
	\caption{\textbf{Statistical moments of $\sigma_{A - B}$ and $\sigma_{A + B}$ as a function of the phase $\phi$ with unitary dynamics -} In the four top
panels, the statistical moments $\langle\sigma^{k}_{A - B}\rangle$ and $\langle\sigma^{k}_{A + B}\rangle$, $k = 1,\ldots,4$, of the stochastic quantum entropy production $\sigma_{A - B}$ and $\sigma_{A + B}$ as a function of $\phi\in[0,2\pi]$ are shown, in the case the dynamics of the composite quantum system $A - B$ is unitary. In the two bottom panels, moreover, we plot a comparison between the samples of the probability distributions $\textrm{Prob}(\sigma_{A - B})$, $\textrm{Prob}(\sigma_{A + B})$ (black squares) and the samples of the corresponding reconstructed distribution (red circles). The latter numerical simulations are performed by considering $\phi = \pi/7$, and $N$ equals, respectively, to $20$ (for the fluctuation profile of $\sigma_{A - B}$) and $10$.}
	\label{fig:fig3}
\end{figure*}
\begin{figure*}[h!]
	\centering
	\includegraphics[scale=6.15]{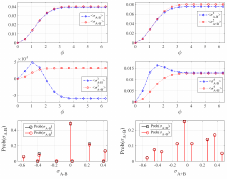}
	\caption{\textbf{Statistical moments of $\sigma_{A - B}$ and $\sigma_{A + B}$ as a function of the phase $\phi$ with noisy dynamics -} In the four top panels, the statistical moments $\langle\sigma^{k}_{A - B}\rangle$ and $\langle\sigma^{k}_{A + B}\rangle$, $k = 1,\ldots,4$, of the stochastic quantum entropy production $\sigma_{A - B}$ and $\sigma_{A + B}$ as a function of $\phi\in[0,2\pi]$ are shown, in the case the dynamics of the composite quantum system $A - B$ is described by a Lindblad (Markovian) equation. In the two bottom panels, moreover, we plot a comparison between the samples of the probability distributions $\textrm{Prob}(\sigma_{A - B})$, $\textrm{Prob}(\sigma_{A + B})$ (black squares) and the samples of the corresponding reconstructed distribution (red circles). The latter numerical simulations are performed by considering $\phi = \frac{5\pi}{6}$, $\Gamma = \Gamma_{A} = \Gamma_{B} = 0.2$ rad/s, and $N$ equals, respectively, to $20$ (for the fluctuation profile of $\sigma_{A - B}$) and $10$.}
	\label{fig:fig4}
\end{figure*}

In Figs.~\ref{fig:fig3} and~\ref{fig:fig4}, we plot the first $4$ statistical moments of $\sigma_{A - B}$ and $\sigma_{A + B}$ as a function of the phase $\phi$, respectively, in case of unitary and noisy dynamics. Moreover, we show, for a given value of $\phi$, the probability distributions $\textrm{Prob}(\sigma_{A - B})$ and $\textrm{Prob}(\sigma_{A + B})$ for both unitary and noisy dynamics, compared with the corresponding reconstructed distributions obtained by applying the reconstruction algorithm, which we call {\small $\overline{\textrm{Prob}(\sigma_{A - B})}$} and {\small $\overline{\textrm{Prob}(\sigma_{A + B})}$}, respectively. Let us recall that $\textrm{Prob}(\sigma_{A + B})$ is obtained by performing the two local measurements with observables $\mathcal{O}^{\textrm{fin}}_{A}$ and $\mathcal{O}^{\textrm{fin}}_{B}$ independently (disregarding the correlations of their outcomes) on the subsystems $A$, $B$, while the distribution $\textrm{Prob}(\sigma_{A - B})$ requires to measure $\mathcal{O}^{\textrm{fin}}_{A}$ and $\mathcal{O}^{\textrm{fin}}_{B}$ simultaneously, {\em i.e.} measuring the observable $\mathcal{O}^{\textrm{fin}}_{A - B}$, defined by Eq.~(\ref{eq:ovservable_A-B}).
For unitary dynamics, the statistical moments of the stochastic quantum entropy productions $\sigma_{A - B}$ and $\sigma_{A + B}$ follow the oscillations of the dynamics induced by changing the gate phase $\phi$. Conversely, for the noisy dynamics of Eq.~(\ref{Lindblad}), with $\Gamma = \Gamma_{A} = \Gamma_{B}$ different from zero, when $\phi$ increases the system approaches a fixed point of the dynamics. Consequently, the statistical moments of the stochastic quantum entropy production tend to the constant values corresponding to the fixed point, and the distribution of the stochastic entropy production becomes narrower. In both Figs.~\ref{fig:fig3} and~\ref{fig:fig4}, the first statistical moments (or mean values) $\langle\sigma_{A - B}\rangle$ and $\langle\sigma_{A + B}\rangle$ are almost overlapping, and the sub-additivity of $\sigma_{A - B}$ is confirmed by the numerical simulations. Furthermore, quite surprisingly, also the second statistical moments of $\sigma_{A - B}$ and $\sigma_{A + B}$ are very similar to each other.
\begin{figure*}[t]
	\centering
	\includegraphics[scale=8.5]{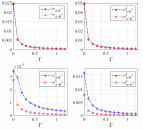}
	\caption{\textbf{Statistical moments of $\sigma_{A-B}$ and $\sigma_{A+B}$ as a function of the dephasing rate $\Gamma$ -} The statistical moments $\langle\sigma^{k}_{A - B}\rangle$ and $\langle\sigma^{k}_{A + B}\rangle$, $k = 1,\ldots,4$, of the stochastic quantum entropy production $\sigma_{A - B}$ and $\sigma_{A + B}$ as a function of $\Gamma\in[0,1.2]$ rad/s are shown, in the case the dynamics of the composite quantum system $A - B$ is described by a Lindblad (Markovian) equation, with $\phi = \pi/7$.}
	\label{fig:fig5}
\end{figure*}
This means that the fluctuation profile of the stochastic entropy production $\sigma_{A + B}$ is able to well reproduce the probability distribution of $\sigma_{A - B}$ in its Gaussian approximation, {\em i.e.} according to the corresponding first and second statistical moments. In addition, we can state that the difference of the higher order moments of $\langle\sigma_{A+B}\rangle$ and $\langle\sigma_{A-B}\rangle$ reflects the presence of correlations between $A$ and $B$ created by the map, since for a product state $\sigma_{A - B} = \sigma_{A+B}$. Therefore, the difference between the fluctuation profiles of $\sigma_{A - B}$ and $\sigma_{A + B}$ constitutes a witness for classical and/or quantum correlations in the final state of the system before the second measurement. As a consequence, if $\textrm{Prob}(\sigma_{A - B})$ and $\textrm{Prob}(\sigma_{A + B})$ are \textit{not} identically equal, then the final density matrix $\rho_{\textrm{fin}}$ is \textit{not} a product state, and (classical and/or quantum) correlations are surely present. Notice that the converse statement is not necessarily true because the quantum correlations can be partially or fully destroyed by the second local measurements, while the classical ones are still preserved and so detectable.

In Fig.~\ref{fig:fig5} the first $4$ statistical moments of $\sigma_{A - B}$ and $\sigma_{A + B}$ are shown as a function of $\Gamma$ (rad/s). As before, we can observe a perfect correspondence between the two quantities when we consider only the first and second statistical moments of the stochastic quantum entropy productions, and, in addition, similar behaviour for the third and fourth statistical moments. Indeed, since the coherence terms of the density matrix describing the dynamics of the composite quantum system tend to zero for increasing $\Gamma$, the number of samples of $\sigma_{A - B}$ and $\sigma_{A + B}$ with an almost zero probability to occur is larger, and also the corresponding probability distribution approaches to a Gaussian one, with zero mean and small variance. In accordance with Figs.~\ref{fig:fig3} and~\ref{fig:fig4}, this result confirms the dominance of decoherence in the quantum system dynamics, which coincides with no creation of correlations.

In the following subsection, we will evaluate the performance of the proposed reconstruction algorithm for the reconstruction of $\textrm{Prob}(\sigma_{A})$ and $\textrm{Prob}(\sigma_{B})$. This choice is largely justified also by the possibility to characterize the irreversibility of an arbitrary quantum process, given by the mean value of the stochastic quantum entropy production $\sigma_{A - B}$, via the reconstruction of the corresponding upper bound in accordance with the sub-additivity property. Still, a similar behaviour was found for the probability distribution $\textrm{Prob}(\sigma_{A-B})$ of the stochastic quantum entropy production of the composite system.

\subsection{Reconstruction for unitary dynamics}

In this section, we show the performance of the reconstruction algorithm for the probability distribution of the stochastic quantum entropy production $\sigma_{A+B}$ via local measurements on the subsystems $A$ and $B$, when the dynamics of the quantum system is unitary. In particular, in the numerical simulations, we take the parameters $\alpha$ and $\beta$ of the algorithm, respectively, equal to the real zeros of the Chebyshev polynomial of degree $N$ in the intervals $\left[\alpha_{\textrm{min}},\alpha_{\textrm{max}}\right] = [0,N]$ and $\left[\beta_{\textrm{min}},\beta_{\textrm{max}}\right] = [0,N]$. This choice for the minimum and maximum values of the parameters $\alpha$ and $\beta$ ensures a very small numerical error (about $10^{-4}$) in the evaluation of each statistical moment of $\sigma_{A}$ and $\sigma_{B}$ via the inversion of the Vandermonde matrix, already for $N > 2$. Indeed, since all the elements of the vectors $\underline{\alpha}$ and $\underline{\beta}$ are different from each other, {\em i.e.} $\alpha_{i}\neq\alpha_{j}$ and $\beta_{i}\neq\beta_{j}$ $\forall i,j = 1,\ldots,N$, we can derive the statistical moments of $\sigma_{C}$, with $C\in\{A,B\}$, by inverting the corresponding Vandermonde matrix. The number $N$ of evaluations of the moment generating functions $\chi_{A}(\alpha)$ and $\chi_{B}(\beta)$, instead, has been taken as a free parameter in the numerics in order to analyze the performance of the reconstruction algorithm. The latter may be quantified in terms of the Root Mean Square Error~(RMSE)
defined as
\begin{equation}\label{RMSE_m}
\textrm{RMSE}\left(\{\langle\sigma^{k}_{A+B}\rangle\}_{k = 1}^{N_{\textrm{max}}}\right)\equiv\sqrt{\frac{\displaystyle{\sum_{k = 1}^{N_{\textrm{max}}}
\left|\langle\sigma^{k}_{A+B}\rangle - \overline{\langle\sigma^{k}_{A+B}\rangle}\right|^{2}}}{N_{\textrm{max}}}},
\end{equation}
where $\{\langle\sigma^{k}_{A+B}\rangle\}$ are the true statistical moments of the stochastic quantum entropy production $\sigma_{A+B}$, which have been numerically computed by directly using Eqs.~(\ref{eq:convolution})-(\ref{prob_b}), while $\overline{\langle\sigma^{k}_{A+B}\rangle}$ are the reconstructed statistical moments after the application of the inverse Fourier transform or the Moore-Penrose pseudo-inverse of $\Sigma_{C}$, $C\in\{A,B\}$. $N_{\textrm{max}}$, instead, is the largest value of $N$ considered for the computation of the $\textrm{RMSE}\left(\{\langle\sigma^{k}_{A+B}\rangle\}\right)$ in the numerical simulations (in this example $N_{\textrm{max}} = 16$).
\begin{figure*}[t]
	\centering
	\includegraphics[scale=6.25]{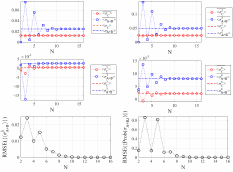}
	\caption{\textbf{Reconstructed statistical moments of $\sigma_{A}$, $\sigma_{B}$ and $\sigma_{A+B}$ as a function of $N$ with unitary dynamics -} In the four top panels we show the statistical moments $\langle\sigma^{k}_{C}\rangle$, $C = \{A,B\}$ (equal due to symmetry), and $\langle\sigma^{k}_{A+B}\rangle$, $k = 1,\ldots,4$, of the stochastic quantum entropy production $\sigma_{A}$, $\sigma_{B}$ and $\sigma_{A+B}$ as a function of $N$. As $N$ increases, the reconstructed statistical moments converge to the corresponding true value. The corresponding RMSEs $\textrm{RMSE}\left(\{\langle\sigma^{k}_{A+B}\rangle\}\right)$ and $\textrm{RMSE}\left(\{\textrm{Prob}(\sigma_{A+B,i})\}\right)$, instead, are plotted in the two bottom panels. All the numerical simulations in the figure are performed by considering unitary dynamics for the composite system $A - B$ with $\phi = \pi/7$.}
	\label{fig:fig6}
\end{figure*}
Another measure for the evaluation of the algorithm performance, which will be used hereafter, is given by the RMSE
\begin{equation}\label{RMSE_p}
\textrm{RMSE}\left(\{\textrm{Prob}(\sigma_{A+B,i})\}_{i = 1}^{l}\right)\equiv\sqrt{\displaystyle{\frac{\sum_{i = 1}^{l}R_{i}^{2}}{l}}},
\end{equation}
where $R_{i} \equiv \left|\textrm{Prob}(\sigma_{A+B,i}) - \overline{\textrm{Prob}(\sigma_{A+B,i})}\right|$ is the reconstruction deviation, {\em i.e.} the discrepancy between the true and the reconstructed probability distribution $\textrm{Prob}(\sigma_{A+B})$. The $\textrm{RMSE}\left(\{\textrm{Prob}(\sigma_{A+B,i})\}\right)$ is computed with respect to the reconstructed values {\small $\overline{\textrm{Prob}(\sigma_{A+B,i})}$} of the probabilities $\textrm{Prob}(\sigma_{A+B,i})$, $i = 1,\ldots,l$, for the stochastic quantum entropy production $\sigma_{A+B}$.

Fig.~\ref{fig:fig6} shows the performance of the reconstruction algorithm as a function of $N$ for the proposed experimental implementation with trapped ions in case the system dynamics undergoes a unitary evolution. In particular, we show the first $4$ statistical moments of $\sigma_{A}$, $\sigma_{B}$ and $\sigma_{A+B}$ as a function of $N$. Let us observe that the statistical moments of the stochastic quantum entropy production of the two subsystems $A$ and $B$ are equal due to the symmetric structure of the composite system. As expected, when $N$ increases, the reconstructed statistical moments converge to the corresponding true values, and also the reconstruction deviation tends to zero. This result is encoded in the RMSEs of Eqs.~(\ref{RMSE_m})-(\ref{RMSE_p}), which behave as monotonically decreasing functions. Both the $\textrm{RMSE}\left(\{\langle\sigma^{k}_{A+B}\rangle\}\right)$ and $\textrm{RMSE}\left(\{\textrm{Prob}(\sigma_{A+B,i})\}\right)$ sharply decrease for about $N \geq 6$, implying that the reconstructed probability distribution {\small $\overline{\textrm{Prob}(\sigma_{A+B})}$} overlaps with the true distribution $\textrm{Prob}(\sigma_{A+B})$ with very small reconstruction deviations $R_{i}$. Since the system of two trapped ions of this example is a small size system, we have chosen to derive the probabilities $\{\textrm{Prob}(\sigma_{A,i})\}$ and $\{\textrm{Prob}(\sigma_{B,i})\}$, $i = 1,\ldots,4$, without performing the inverse Fourier transform on the statistical moments $\{\widetilde{\langle\sigma_{C}^{k}\rangle}\}$, $C\in\{A,B\}$. Indeed, the computation of the inverse Fourier transform, which has to be performed numerically, can be a tricky step of the reconstruction procedure, because it can require the adoption of numerical methods with an adaptive step-size in order to solve the numerical integration. In this way, the only source of error in the reconstruction procedure is given by the expansion in Taylor series of the quantity $\chi_{C}(\varphi)$, with $C\in\{A,B\}$ and $\varphi\in\{\alpha,\beta\}$, around $\varphi = 0$ as a function of a \textit{finite} number of statistical moments $\langle\sigma_{C}^{k}\rangle$, $k = 1,\ldots,N-1$. As shown in Fig.~\ref{fig:fig6}, the choice of the value of $N$ is a degree of freedom of the algorithm, and it strictly depends on the physical implementation of the reconstruction protocol. In the experimental implementation above with two trapped ions, $N = 10$ ensures very good performance without making a larger number of measurements with respect to the number of values assumed by the stochastic quantum entropy production $\sigma_{A+B}$.
\begin{figure*}[h!]
	\centering
	\includegraphics[scale = 5.5]{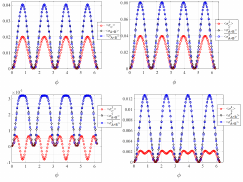}
	\caption{\textbf{True and reconstructed statistical moments of $\sigma_{A}$, $\sigma_{B}$ and $\sigma_{A + B}$ as a function of the phase $\phi$ with unitary dynamics -} The statistical moments $\langle\sigma^{k}_{C}\rangle$, $C = \{A,B\}$ (equal by symmetry), and $\langle\sigma^{k}_{A + B}\rangle$, $k = 1,\ldots,4$, of the stochastic quantum entropy production $\sigma_{A}$, $\sigma_{B}$ and $\sigma_{A + B}$ are shown as a function of the phase $\phi$. All the numerical simulations are performed by considering unitary dynamics for the composite system $A - B$ with $N = 10$ and $\phi\in[0,2\pi]$.}
	\label{fig:Fig7}
\end{figure*}

In Fig.~\ref{fig:Fig7}, moreover, we show for $N = 10$ the first $4$ true statistical moments of the stochastic quantum entropy productions $\sigma_{A}$ and $\sigma_{B}$ of the two subsystems, as well as the correlation-free convolution $\sigma_{A+B}$ as a function of $\phi\in[0,2\pi]$, along with the corresponding reconstructed counterpart {\small $\overline{\langle\sigma^{k}_C\rangle}$}, $k = 1,\ldots,4$, $C\in\{A,B,A+B\}$. As before, the reconstruction procedure yields values very close to the true statistical moments of $\sigma_{A}$, $\sigma_{B}$ and $\sigma_{A+B}$ for all values of the phase $\phi$.

\subsection{Reconstruction for noisy dynamics}

Here, the performance of the reconstruction algorithm is discussed in case the system dynamics is affected by pure-dephasing contributions, described via the differential Lindblad (Markovian) equation $\dot{\rho}(t) = \mathcal{L}(\rho(t))$, given by Eq.~(\ref{Lindblad}). The Hamiltonian of the composite system $A - B$ is defined as $H = \omega\left(X^{A}\otimes X^{B}\right)$. Since the fixed duration $\tau$ of the transformation has been chosen as before equal to $50$ s, we choose the desired phase $\phi$ by setting the interaction strength to $\omega \equiv \phi/\tau$ (rad/s).
\begin{figure}[h!]
	\centering
	\includegraphics[scale = 6]{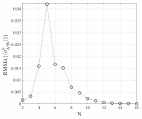}
	\caption{\textbf{RMSE from the reconstructed statistical moments as a function of $N$ -} Root mean square error $\textrm{RMSE}\left(\{\langle\sigma^{k}_{A+B}\rangle\}\right)$ as a function of $N$. All the numerical simulations are performed by considering a Lindblad (Markovian) dynamics for the composite system $A - B$, given by Eq.~(\ref{Lindblad}), where the dephasing rates $\Gamma = \Gamma_{A} = \Gamma_{B}$ are chosen equal to $0.2$ rad/s, and $\phi = \pi/7$.}
	\label{fig:entropy_Lindblad}
\end{figure}

\begin{figure*}[h!]
	\centering
	\includegraphics[scale = 6.25]{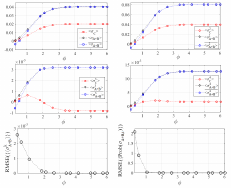}
	\caption{\textbf{True and reconstructed statistical moments of $\sigma_{A}$, $\sigma_{B}$ and $\sigma_{A+B}$ as a function of the phase $\phi$ with noisy dynamics -} In the first $4$ panels the statistical moments $\langle\sigma^{k}_{C}\rangle$, $C = \{A,B\}$ (equal by symmetry), and $\langle\sigma^{k}_{A+B}\rangle$, $k = 1,\ldots,4$, of the stochastic quantum entropy production $\sigma_{A}$, $\sigma_{B}$ and $\sigma_{A+B}$ are shown as a function of the phase $\phi$. All the numerical simulations are performed by considering a Lindblad (Markovian) dynamics for the composite system $A - B$, given by Eq.~(\ref{Lindblad}), with $N = 10$, $\Gamma = 0.2$ rad/s, and $\phi\in[0,2\pi]$. In the bottom panels of the figure, instead, we show the root mean square errors $\textrm{RMSE}\left(\{\langle\sigma^{k}_{A+B}\rangle\}\right)$ and $\textrm{RMSE}\left(\{\textrm{Prob}(\sigma_{A+B,i})\}\right)$.}
	\label{fig:Fig9}
\end{figure*}

Fig~\ref{fig:entropy_Lindblad} shows the $\textrm{RMSE}\left(\{\langle\sigma^{k}_{A+B}\rangle\}\right)$ computed from the reconstructed statistical moments of the stochastic quantum entropy production $\sigma_{A+B}$. As it can be observed, apart from an initial transient, the mean error monotonically tends to zero as $N$ increases, similarly to the case of unitary dynamics (see Fig.~\ref{fig:fig6}), such that it can be considered sufficiently small for $N > 8$. Again, we evaluate the performance of the reconstruction algorithm also as a function of the phase $\phi = \omega\tau$, with $\tau$ fixed. As shown in Fig.~\ref{fig:Fig9}, when $\phi$ increases (with a fixed value of $\Gamma$, set to $0.2$ rad/s) the statistical moments of $\sigma_{A+B}$ (but not necessarily the ones regarding the subsystems $A$ and $B$) increase as well, since when $\phi$ increases the system tends to a fixed point of the dynamics. Also the reconstruction procedure turns out to be more accurate for larger values of $\phi$, as shown in the two bottom panels of Fig.~\ref{fig:Fig9} (for this figure we use the Fourier transform). The reason is that when the dynamics approaches the fixed point, the distribution of the stochastic quantum entropy production becomes narrower and the convergence of the Fourier integral is ensured.

\begin{figure*}[h!]
	\centering
	\includegraphics[scale = 8.5]{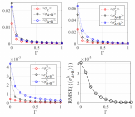}
	\caption{\textbf{True and reconstructed statistical moments of $\sigma_{A}$, $\sigma_{B}$ and $\sigma_{A+B}$ as a function of the dephasing rate $\Gamma$ -} The first $3$ statistical moments of the stochastic quantum entropy productions for $A$, $B$ (equal by symmetry) and the composite system $A - B$ as a function of the dephasing rate $\Gamma = \Gamma_{A} = \Gamma_{B}$ (rad/s) are shown for the physical example of $2$ trapped ions. In particular, the statistical moments of $\sigma_{A+B}$ are put beside their reconstructed version, which have been obtained by choosing $N = 10$ and $\phi = \pi/7$. In the last panel, moreover, the corresponding root mean square error $\textrm{RMSE}\left(\{\langle\sigma^{k}_{A+B}\rangle\}\right)$ as a function of $\Gamma$ is shown.}
	\label{fig:Fig10}
\end{figure*}

Finally, in Fig.\ref{fig:Fig10} we plot the behaviour of the first three statistical moments of $\sigma_{A}$, $\sigma_{B}$ and $\sigma_{A+B}$ as a function of the dephasing rate $\Gamma = \Gamma_{A} = \Gamma_{B}$ (rad/s), with $N = 10$ and $\phi = \pi/7$. As before, due to the symmetry of the bipartition, the statistical moments of $\sigma_{A}$ and $\sigma_{B}$ are identically equal. For increasing $\Gamma$ the dephasing becomes dominant over the interaction and all correlations between the subsystems are destroyed. As a consequence, the stochastic quantum entropy production tends to zero as is observed in the figure for all the investigated moments, both for the subsystems and the composite system.

\subsection{Probing irreversibility}

Once the fluctuation profile of the stochastic quantum entropy production ({\em i.e.} the corresponding probability distribution) is reconstructed, then the irreversibility properties of the composite system transformation can be successfully probed. The thermodynamic irreversibility, indeed, is quantified by means of the mean value $\langle\sigma_{A - B}\rangle$, with $\langle \sigma_{A - B} \rangle = 0$ corresponding to thermodynamic reversibility. As previously shown in Figs.~\ref{fig:fig3},~\ref{fig:fig4} and \ref{fig:fig5}, the mean value $\langle\sigma_{A - B}\rangle$ can be well approximated by $\langle\sigma_{A + B}\rangle$ and from Eqs.~(\ref{eq:entropy-positivity}) and (\ref{eq:sigma_sum}) we have $0\leq\langle\sigma_{A-B}\rangle \leq \langle\sigma_{A+B}\rangle$. From Figs.~\ref{fig:fig5} and~\ref{fig:Fig10}, thus, we can observe that the implemented noisy transformation is more reversible with respect to the unitary one. Indeed, the statistical moments of the stochastic quantum entropy production, as well as the corresponding mean value, go to zero as $\Gamma$ increases. Since the dynamics originating from the Lindblad equation~(\ref{Lindblad}) admits as a fixed point the completely mixed state of the composite system $A - B$, if we increase the value of $\Gamma$ then the probability distribution of the quantum entropy production for the systems $A$, $B$ and $A - B$ tends to a Kronecker delta around zero, leading to a more-reversible system transformation with respect to the unitary case. For this reason, also the numerical simulations of Fig.~\ref{fig:Fig10} have been performed by using the inverse Fourier transform to reconstruct the probabilities $\{\textrm{Prob}(\sigma_{C,i})\}$, with $i = 1,\ldots,M_{C}$ and $C\in\{A,B\}$, instead of calculating the pseudo-inverse of the matrix $\Sigma_{C}$. As a matter of fact, as $\Gamma$ increases some values of $\sigma_{C}$ approach zero and $\Sigma_{C}$ becomes singular. Let us observe that when the dynamics is unitary the performance of the reconstruction algorithm adopting the inverse Fourier transform can be affected by a not-negligible error, as shown by the $\textrm{RMSE}\left(\{\langle\sigma^{k}_{A+B}\rangle\}\right)$ in the last panel of Fig.\ref{fig:Fig10}. For such case, the adoption of the pseudo-inverse of $\Sigma_{C}$ is to be preferred. Moreover, we expect that increasing the number of ions the thermodynamic irreversibility becomes more and more pronounced and that
this analysis may be the object of forthcoming work.

In conclusion, a system transformation on a multipartite quantum system involves stochastic quantum entropy production whenever correlations between the subsystems of the multipartite system is first created by the dynamics of the composite system and then destroyed by the second measurement. This result, indeed, can be easily deduced from Figs.~\ref{fig:Fig9} and~\ref{fig:Fig10}, in which, for a fixed value of $\Gamma$ ($\Gamma$ = 0.2 rad/s) and $\phi$ $(\phi = \pi/7)$ respectively, the behaviour of the statistical moments of the stochastic quantum entropy production as a function of $\phi$ ($\Gamma$) is monotonically increasing (decreasing). Indeed, the stronger is the interaction between the two ions, the larger is the corresponding production of correlations between them. On the other side, instead, the production of correlations within a multipartite system is inhibited due to the presence of strong decoherent processes.

\section{Discussion and conclusions}

The goal of this paper has been three-fold:

We discussed how to relate the stochastic quantum entropy production to the quantum fluctuation theorem, which is the generalization of the Tasaki-Crooks theorem for open systems, and this relation is based on the evaluation and quantification of the irreversibility by means of a two-times measurement protocol. By using the definition of the stochastic quantum entropy production, then, we introduced a protocol to reconstruct its fluctuations, and characterize the dynamics of the quantum system. In particular, the proposed procedure allows us to determine the mean value $\langle\sigma\rangle$ of the stochastic quantum entropy production, quantifying and probing the amount of irreversibility, with $\langle \sigma \rangle=0$ ($\langle \sigma \rangle>0$) corresponding to thermodynamic reversibility (irreversibility). At variance, $\langle \sigma \rangle<0$ -- violating the second law of the thermodynamics -- do not occur due to the non-negativity of the Kullback-Leibler divergence. Furthermore, we have shown the relation between the quantum relative entropy of the system density matrix at the final time of the transformation and the stochastic quantum entropy production. In this regard, Eq.~(\ref{eq:entropy-positivity}) is valid also at zero temperature and without assuming thermal baths with finite temperature. It is worth noting, however, that the presented results about the stochastic quantum entropy production are based on the assumption to consider unital CPTP quantum maps for the description of the open system dynamics. Even though we limited ourself for most of the paper to unital CPTP quantum maps, in Section \ref{rec_2nd_law} we discussed how to put in relation our results with the conventional second law of thermodynamics at finite temperature.

Secondly, to reconstruct the probability distribution of $\sigma$, we proposed a suitable reconstruction protocol based on the determination of the characteristic functions of the stochastic quantum entropy production, which are evaluated $N$ times for a given set of $N$ real parameters, so as to collect an adequate amount of information for the reconstruction. In other words, the reconstruction algorithm relies on a parametric version of the integral quantum fluctuation theorem, and yields the first $N$ statistical moments of the stochastic quantum entropy production through the inversion of a Vandermonde matrix, which encodes the experimental evaluation of the characteristic function. The corresponding numerical error, however, can be reduced using the solution of a polynomial interpolation problem based on the use of (zeros of) the Chebyshev polynomials. Then, the presented algorithm in its generic form uses a numerical inverse Fourier transform to reconstruct the probability distribution of the stochastic quantum entropy production from its statistical moments. In order to optimize the minimum amount of resources needed for the protocol, we have also shown that the required number of measurements to perform the algorithm scales linearly with the number of the values that can be assumed by the stochastic quantum entropy production, and not quadratically as one would have obtained by a direct application of the definition of the corresponding probability distributions.

Moreover, as third aspect, under the hypothesis that the open quantum system is composed by mutually interacting subsystems, we investigated the stochastic quantum entropy production both for the subsystems and for the composite system, showing that their mean values are sub-additive; they coincide only when the system density matrix at the end of the transformation is a product state. In this way, we have proved not only how to characterize the irreversibility of an arbitrary unital quantum process by reconstructing the corresponding upper-bound in accordance with the sub-additivity, but we have also provided a method to detect correlations between the partitions of the system, just by comparing the fluctuation profiles of $\sigma_{A+B}$ and $\sigma_{A-B}$. In this regard, let us recall that the correlation shared by two parts of a composite system (let us consider e.g. even just the entanglement, genuinely quantum type of correlation) is not directly measurable in laboratory, since there does not exist a self-adjoint operator quantifying it. To this aim, one could perform a single quantum state tomography on the final state of the system after the transformation and evaluate the quantum discord for such state, so as to measure nonclassical correlations between its partitions \cite{OllivierPRL2001}. However, it is worth noting that the performance of a full state tomography is not always feasible in terms of the available resources, and, along this direction, nontrivial solutions have been recently introduced (see e.g. Ref. \cite{GirolamiPRL2012}). Conversely, as previously shown, the measure of the probability distribution of the quantum entropy production by performing measurements both on local and global observables of the composite system does not require any quantum state tomography, since only measurements of the occupation probabilities of the final density matrix of the system are needed. Finally, to detect correlations one could also directly measure the second order dynamical correlation function of the system. However, such procedure, although it could not require a full quantum state tomography, is extremely system-dependent and usually a relevant experimental effort is necessary, though it has been recently introduced a measurement apparatus relying on weak-ancilla-system couplings, which could in part facilitate such measurements \cite{UhrichPRA2017}.

In order to illustrate our theoretical results, we discussed an experimental implementation with trapped ions and we showed the performance of the reconstruction algorithm on a quantum system composed of two trapped ions, subjected to a unitary evolution and to a Lindbladian one. We showed that the probability distribution of the stochastic quantum entropy production can be efficiently reconstructed with a very small error already with $N = 7$ momenta. Efficiency and possible extensions of the presented procedure have been also discussed. Generalizations of our results to more general (non-unital) quantum maps have been briefly discussed and will be investigated in a future work. Our protocol, summarized in Fig.~\ref{fig:procedure}, is experimentally oriented and it is based on the preparation of suitably prepared initial states, depending on the parameter $\varphi$ entering the characteristic functions $\chi_{C}(\varphi)$ to be measured. Such protocol appears to be within the reach of experimental realizations, given the remarkable results obtained in the last decades in the preparation of quantum states~\cite{BergmannRMP1998}. In this regard, we observe that state preparation can be achieved in most common quantum technology platforms via optimal control procedures~\cite{BrifNJP2010}, allowing to speed up the process of coherent population transfer up to the ultimate bound imposed by quantum mechanics, {\em i.e.} the quantum speed limit~\cite{CanevaPRL2009,DoriaPRL2011}. As a final remark, we also observe that the fluctuation properties of the stochastic quantum entropy production do strongly depend on the presence of decoherent channels between an arbitrary quantum system and the environment. Hence, one could effectively determine not only the influence of the external noise sources on the system dynamics, but also characterize the environment structure and properties via quantum sensing procedures. We believe that this aspect could be the subject for future investigations, {\em e.g.} along the research direction proposed in~\cite{Mueller3,Mueller2}, based on stochastic quantum Zeno phenomena~\cite{Gherardini1,Mueller1}, and/or in Refs.~\cite{Cosco2015,Nokkala2016} via engineered quantum networks. \\ \\
{\bf Acknowledgments} \\
The authors gratefully acknowledge Mauro Paternostro, Giorgio Battistelli, Giacomo Gori, Jin Wang, Francesco Saverio Cataliotti, Duccio Fanelli, Pietro Silvi and Augusto Smerzi for useful discussions. S.G. and M.M.M. thank the Scuola Internazionale Superiore di Studi Avanzati (SISSA), Trieste (Italy) for hospitality during the completion of this work. A.T. thanks the Galileo Galilei Institute for Theoretical Physics for the hospitality in the Workshop ``From Static to Dynamical Gauge Fields with Ultracold Atoms'' and the INFN for partial support during the completion of this work. This work was financially supported from the Fondazione CR Firenze through the project Q-BIOSCAN.

\appendix

\section{Proof of Theorem~1}\label{appendix_a}

In this Appendix, we prove the equality between the conditional probabilities $p(a^{\textrm{fin}}_{k}|a^{\textrm{in}}_{m})$ and $p(a^{\textrm{in}}_{m}|a^{\textrm{ref}}_{k})$, respectively, of the forward and backward processes of our two-time measurement scheme.
Let us recall the observables $\mathcal{O}_{\textrm{in}}\equiv\sum_{m}a^{\textrm{in}}_{m}\Pi^{\textrm{in}}_{m}$, $\mathcal{O}_{\textrm{fin}}\equiv\sum_{k}a^{\textrm{fin}}_{k}\Pi^{\textrm{fin}}_{k}$, $\widetilde{\mathcal{O}}_{\textrm{ref}}\equiv\sum_{k}a^{\textrm{ref}}_{k}\widetilde{\Pi}^{\textrm{ref}}_{k}$ and $\widetilde{\mathcal{O}}_{\textrm{in}} = \sum_{m}a^{\textrm{in}}_{m}\widetilde{\Pi}^{\textrm{in}}_{m}$, as defined in the main text. The dynamical evolution of the open quantum system between the two measurements is described by a unital CPTP map $\Phi(\cdot)$ (with $\Phi(\mathbbm{1}) = \mathbbm{1}$), whose Kraus operators $\{E_{u}\}$ are such that $\sum_{u}E_{u}^{\dagger}E_{u} = \mathbbm{1}$, where $\mathbbm{1}$ denotes the identity operator on the Hilbert space $\mathcal{H}$ of the quantum system.
Accordingly, $\Phi(\rho_{\textrm{in},m}) = \sum_{u}E_{u}\rho_{\textrm{in},m}E_{u}^{\dagger}$, where $\rho_{\textrm{in},m}\equiv\Pi^{\textrm{in}}_{m}\rho_{0}\Pi^{\textrm{in}}_{m}$, and thus the conditional probability $p(a^{\textrm{fin}}_{k}|a^{\textrm{in}}_{m})$ equals to
\begin{eqnarray}\label{cond_prob_forward_appA}
&&p(a^{\textrm{fin}}_{k}|a^{\textrm{in}}_{m}) = \frac{\textrm{Tr}[\Pi^{\textrm{fin}}_{k}\Phi(\rho_{\textrm{in},m})]}
{\textrm{Tr}[\Pi^{\textrm{in}}_{m}\rho_{0}\Pi^{\textrm{in}}_{m}]} = \frac{\textrm{Tr}[\Pi^{\textrm{fin}}_{k}\sum_{u}E_{u}\rho_{\textrm{in},m}E_{u}^{\dagger}]}
{\textrm{Tr}[\Pi^{\textrm{in}}_{m}\rho_{0}\Pi^{\textrm{in}}_{m}]}\nonumber \\
&&= \sum_{u}\frac{\textrm{Tr}[\Pi^{\textrm{fin}}_{k}E_{u}\Pi^{\textrm{in}}_{m}\rho_{0}\Pi^{\textrm{in}}_{m}E_{u}^{\dagger}]}
{\textrm{Tr}[\Pi^{\textrm{in}}_{m}\rho_{0}\Pi^{\textrm{in}}_{m}]} = \sum_{u}|\langle\phi_{a_{k}}|E_{u}|\psi_{a_{m}}\rangle|^{2}.\nonumber \\
&&
\end{eqnarray}
Next, by inserting in Eq.~(\ref{cond_prob_forward_appA}) the identity operator $\mathbbm{1} = \Theta\Theta^\dagger=\Theta^\dagger\Theta$, where $\Theta$ is the time-reversal operator as defined in the main text, one has:
\begin{eqnarray}
|\langle\phi_{a_{k}}|E_{u}|\psi_{a_{m}}\rangle|^{2} &=& |\langle\phi_{a_{k}}|\Theta^{\dagger}\left(\Theta E_{u}\Theta^\dagger \right)\Theta|\psi_{a_{m}}\rangle|^{2}
= |\langle \widetilde{\phi}_{a_{k}}| \Theta E_{u} \Theta^\dagger |\widetilde{\psi}_{a_{m}}\rangle|^2 \nonumber\\
&=& |\langle\widetilde{\psi}_{a_{m}}|\Theta E_{u}^{\dagger}\Theta^\dagger|\widetilde{\phi}_{a_{k}}\rangle|^{2}.
\end{eqnarray}
where we have used complex conjugation and the modulus squared to flip the order of the operators. The time-reversal of a single Kraus operator is $\widetilde{E}_{u}\equiv\mathcal{A}\pi^{1/2}E^{\dagger}_{u}\pi^{-1/2}\mathcal{A}^{\dagger}$, where we choose $\mathcal{A} = \Theta$ and $\pi = \mathbbm{1}$ (as $\Phi$ is unital, such that $\Phi(\mathbbm{1}) = \mathbbm{1}$). We can now state that
\begin{equation}
|\langle\phi_{a_{k}}|E_{u}|\psi_{a_{m}}\rangle|^{2} = |\langle\widetilde{\psi}_{a_{m}}|\widetilde{E}_{u}|\widetilde{\phi}_{a_{k}}\rangle|^{2}.
\end{equation}
Moreover, by observing that
\begin{equation}
\sum_{u}|\langle\widetilde{\psi}_{a_{m}}|\widetilde{E}_{u}|\widetilde{\phi}_{a_{k}}\rangle|^{2} = \frac{\textrm{Tr}[\widetilde{\Pi}^{\textrm{in}}_{m}\widetilde{\Phi}(\rho_{\textrm{ref},k})]}
{\textrm{Tr}[\widetilde{\Pi}^{\textrm{ref}}_{k}\widetilde{\rho}_{\tau}\widetilde{\Pi}^{\textrm{ref}}_{m}]}
= p(a^{\textrm{in}}_{m}|a^{\textrm{ref}}_{k}),
\end{equation}
where $\rho_{\textrm{ref},k}\equiv\widetilde{\Pi}^{\textrm{ref}}_{k}\widetilde{\rho}_{\tau}\widetilde{\Pi}^{\textrm{ref}}_{m}$, the equality $p(a^{\textrm{fin}}_{k}|a^{\textrm{in}}_{m}) = p(a^{\textrm{in}}_{m}|a^{\textrm{ref}}_{k})$, as well as the Theorem~1, follow straightforwardly.

\section{Proof of Theorem~2}\label{appendix_b}

Here, we prove Theorem~2, {\em i.e.} the inequality
\begin{equation*}
0\leq S(\rho_{\textrm{fin}}\parallel\rho_{\tau})\leq\langle\sigma\rangle,
\end{equation*}
where $\rho_{\textrm{fin}}$ and $\rho_{\tau}$ are the density operators of the open quantum system $\mathcal{S}$ before and after the second measurement of the forward process. $S(\rho_{\textrm{fin}}\parallel\rho_{\tau})$ is called the quantum relative entropy of $\rho_{\textrm{fin}}$ and $\rho_{\tau}$ and $\langle\sigma\rangle$ is the average of the stochastic quantum entropy production. This inequality may be regarded as the quantum counterpart of the second law of thermodynamics for an open quantum system.

Let us consider the stochastic entropy production $\sigma(a^{\textrm{fin}},a^{\textrm{in}})=\ln\left[\frac{p(a^{\textrm{in}})}{p(a^{\textrm{ref}})}\right]$ (as given in Eq.~(\ref{sigma}) in the main text) for the open quantum system $\mathcal{S}$, whose validity is subordinated to the assumptions of Theorem~1. Accordingly, the average value of $\sigma$ is
\begin{equation}
\langle\sigma\rangle = \sum_{a^{\textrm{fin}},a^{\textrm{in}}}p(a^{\textrm{fin}},a^{\textrm{in}})\ln\left[\frac{p(a^{\textrm{in}})}{p(a^{\textrm{ref}})}\right]
= \sum_{a^{\textrm{in}}}p(a^{\textrm{in}})\ln[p(a^{\textrm{in}})] - \sum_{a^{\textrm{fin}}}p(a^{\textrm{fin}})\ln[p(a^{\textrm{ref}})]\geq 0.
\end{equation}
We observe that the mean quantum entropy production $\langle\sigma\rangle$ is a non-negative quantity due to the positivity of the classical relative entropy, or Kullback-Leibler divergence. Since $p(a^{\textrm{fin}})\equiv\langle\phi_{a}|\rho_{\textrm{fin}}|\phi_{a}\rangle$ and the reference state is diagonal in the basis $\{|\phi_a\rangle\}$, we have
\begin{equation}
\sum_{a^{\textrm{fin}}}p(a^{\textrm{fin}})\ln[p(a^{\textrm{ref}})] = \sum_{a^{\textrm{fin}}}\langle\phi_{a}|\rho_{\textrm{fin}}|\phi_{a}\rangle\ln[p(a^{\textrm{ref}})]
= \sum_{a^{\textrm{fin}}}\langle\phi_{a}|\rho_{\textrm{fin}}\ln\rho_{\textrm{ref}}|\phi_{a}\rangle =
\textrm{Tr}\left[\rho_{\textrm{fin}}\ln\rho_{\tau}\right],
\end{equation}
where the last identity is verified by assuming the equality between the reference state $\rho_{\textrm{ref}}$ and the density operator $\rho_{\tau}$ after the second measurement of the protocol. One also has:
\begin{equation}
\sum_{a^{\textrm{in}}}p(a^{\textrm{in}})\ln[p(a^{\textrm{in}})] = \textrm{Tr}\left[\rho_{\textrm{in}}\ln\rho_{\textrm{in}}\right] = -S(\rho_{\textrm{in}}),
\end{equation}
where $S(\rho_{\textrm{in}})\equiv-\textrm{Tr}\left[\rho_{\textrm{in}}\ln\rho_{\textrm{in}}\right]$ is the von Neumann entropy for the initial density operator $\rho_{\textrm{in}}$ of the quantum system $\mathcal{S}$. The mean quantum entropy production $\langle\sigma\rangle$, thus, can be written in general as
\begin{equation}
\langle\sigma\rangle = -\textrm{Tr}\left[\rho_{\textrm{fin}}\ln\rho_{\tau}\right] - S(\rho_{\textrm{in}}).
\end{equation}
The quantum relative entropy is defined as $S(\rho_{\textrm{fin}}\parallel\rho_{\tau})=- \textrm{Tr}\left[\rho_{\textrm{fin}}\ln\rho_{\tau}\right] - S(\rho_{\textrm{fin}})$
and trivially $S(\rho_{\textrm{fin}}\parallel\rho_{\tau})\geq 0$. According to our protocol, the initial and the final states are connected by the unital CPTP map $\Phi$ as $\rho_{\textrm{fin}}=\Phi(\rho_{\textrm{in}})$. As a consequence of the unitality of $\Phi$
the von Neumann entropies obey the relation
$S(\rho_{\textrm{in}})\leq S(\rho_{\textrm{fin}})$~\cite{Sagawa2014}.
Summarizing, we obtain
\begin{equation}
0\leq S(\rho_{\textrm{fin}}\parallel\rho_{\tau}) = - \textrm{Tr}\left[\rho_{\textrm{fin}}\ln\rho_{\tau}\right] - S(\rho_{\textrm{fin}})
\leq - \textrm{Tr}\left[\rho_{\textrm{fin}}\ln\rho_{\tau}\right] - S(\rho_{\textrm{in}}) = \langle\sigma\rangle,
\end{equation}
proving the original inequality.

Note that if we perform the second measurement with a basis in which $\rho_{\textrm{fin}}$ is diagonal ({\em i.e.} vanishing commutator between measurement operator and final state, $[\mathcal{O}_{\textrm{fin}},\rho_{\textrm{fin}}]=0$), the state is unchanged by the second measurement and $\rho_{\textrm{fin}}=\rho_{\tau}$. As a consequence
\begin{equation*}
0= S(\rho_{\textrm{fin}}\parallel\rho_{\tau})\leq\langle\sigma\rangle=S(\rho_{\textrm{fin}})-S(\rho_{\textrm{in}}),
\end{equation*}
{\em i.e.} the quantum relative entropy vanishes, while the average of the stochastic entropy production equals to the difference of final and initial von Neumann entropies, $\langle\sigma\rangle=S(\rho_{\textrm{fin}})-S(\rho_{\textrm{in}})$, and thus describes the irreversibility distribution of the map $\Phi$ only (and not of the measurement, as it would be in the general case).

In the general case, i.e. if the condition $[\mathcal{O}_{\textrm{fin}},\rho_{\textrm{fin}}]=0$ does not hold, still the post-measurement state $\rho_{\tau}$ is diagonal in the basis of the observable eigenstates and we obtain
\begin{equation}
\langle\sigma\rangle = -\textrm{Tr}\left[\rho_{\textrm{fin}}\ln\rho_{\tau}\right] - S(\rho_{\textrm{in}}) = S(\rho_{\tau})- S(\rho_{\textrm{in}}).
\end{equation}

\section{Recovering the second law of thermodynamics}
\label{appendix_second_law}

Here we show the formal derivation to connect the entropy production inequality in Theorem 2 to the second law of thermodynamics for the open quantum system $\mathcal{S}$ in term of the mean work $\langle\mathrm{W}\rangle$ and the free-energy difference $\Delta F$ as defined in the main text. The main ingredient of this proof is to express ${\rm Tr}[\rho_{\textrm{fin}}\ln\rho_{\tau}]$ as a function of the thermal state $\rho_{\tau}^{{\rm th}} \equiv e^{\beta\left[F(\tau)\mathbbm{1}_\mathcal{S} - H(\tau)\right]}$ at time $t = \tau$. In particular, we can write that
\begin{eqnarray*}
{\rm Tr}[\rho_{\textrm{fin}}\ln\rho_{\tau}] &=& {\rm Tr}[\rho_{\textrm{fin}}\ln\rho_{\tau}^{{\rm th}}] + {\rm Tr}[\rho_{\textrm{fin}}\ln\rho_{\tau}] - {\rm Tr}[\rho_{\textrm{fin}}\ln\rho_{\tau}^{{\rm th}}] \\
&=& {\rm Tr}[\rho_{\textrm{fin}}\ln\rho_{\tau}^{{\rm th}}] + {\rm Tr}[\rho_{\textrm{fin}}(\ln\rho_{\textrm{fin}} - \ln\rho_{\textrm{fin}} + \ln\rho_{\tau} - \ln\rho_{\tau}^{{\rm th}})]                      \\
&=& {\rm Tr}[\rho_{\textrm{fin}}\ln\rho_{\tau}^{{\rm th}}] + {\rm Tr}[\rho_{\textrm{fin}}(\ln\rho_{\textrm{fin}} - \ln\rho_{\tau}^{{\rm th}})] - {\rm Tr}[\rho_{\textrm{fin}}(\ln\rho_{\textrm{fin}} - \ln\rho_{\tau})], \\
\end{eqnarray*}
so as to obtain
\begin{equation}
{\rm Tr}[\rho_{\textrm{fin}}\ln\rho_{\tau}] = {\rm Tr}[\rho_{\textrm{fin}}\ln\rho_{\tau}^{{\rm th}}] + S(\rho_{\textrm{fin}}\parallel\rho_{\tau}^{{\rm th}}) - S(\rho_{\textrm{fin}}\parallel\rho_{\tau}),
\end{equation}
i.e. Eq. (\ref{new_equation_quantum_entropy}) in the main text. Therefore, by taking Eq. (\ref{sigma_theorem2}) and substituting
\begin{equation*}
  S(\rho_{\textrm{in}}) = - {\rm Tr}[\rho_{\textrm{in}}\ln\rho_{\textrm{in}}] = -\beta F(0) + \beta{\rm Tr}[\rho_{\textrm{in}}H(0)]
\end{equation*}
with $\rho_{\textrm{in}} \equiv e^{\beta\left[F(0)\mathbbm{1}_\mathcal{S} - H(0)\right]}$, one has
\begin{equation*}
  \langle\sigma\rangle = \beta F(0) - \beta{\rm Tr}[\rho_{\textrm{in}}H(0)] - \beta F(\tau) + \beta{\rm Tr}[\rho_{\textrm{fin}}H(\tau)] - S(\rho_{\textrm{fin}}\parallel\rho_{\tau}^{{\rm th}}) + S(\rho_{\textrm{fin}}\parallel\rho_{\tau}).
\end{equation*}
Accordingly, being $0\leq S(\rho_{\textrm{fin}}\parallel\rho_{\tau})\leq\langle\sigma\rangle$ (from the results of Theorems 1 and 2) and $S(\rho_{\textrm{fin}}\parallel\rho_{\tau}^{{\rm th}})\geq 0$ (non-negativity of the quantum relative entropy), we finally recover the conventional second law of thermodynamics, i.e. $\langle\mathrm{W}\rangle\geq\Delta F$.

\section{Derivation of the characteristic functions}\label{appendix_charact_func}

In this Appendix, we derive the expressions for the characteristic functions $G_A(\lambda)$ (for the probability distributions $\textrm{Prob}(\sigma_A)$) and $G_B(\lambda)$ (for the probability distributions $\textrm{Prob}(\sigma_B)$), given by Eq.~(\ref{eq3}) and Eq.~(\ref{eq4}), respectively.
We start with the definition
\begin{equation}
G_A(\lambda) = \int \textrm{Prob}_{A}(\sigma_A)e^{i\lambda\sigma_A}d\sigma_A,
\end{equation}
where
\begin{equation}
\textrm{Prob}(\sigma_A) = \sum_{k,m}\delta\left[\sigma_A - \sigma_A(a^{\textrm{in}}_{m},a^{\textrm{fin}}_{k})\right]p_a(k,m),
\end{equation}
as well as
\begin{equation}
p_{a}(k,m) = \textrm{Tr}\left[(\Pi^{\tau}_{A,k}\otimes\mathbbm{1}_{B})\Phi(\Pi^{\textrm{in}}_{A,m}\otimes\rho_{\textrm{B,in}})\right]p(a_{m}^{\textrm{in}}),
\end{equation}
and
\begin{equation}
\sigma_A(a^{\textrm{in}}_{m},a^{\textrm{fin}}_{k})=\ln [p(a^{\textrm{in}}_{m})] - \ln[p(a^{\textrm{fin}}_{k})]
\end{equation}
Exploiting the linearity of $\Phi$ and the trace, we obtain
\begin{eqnarray}\label{eq:G-appendix}
&&G_A(\lambda) = \sum_{k,m}p_{a}(k,m)e^{i\lambda\sigma_A(a^{\textrm{in}}_{m},a^{\textrm{fin}}_{k})}\nonumber \\
&&= \textrm{Tr}\left[\left(\sum_{k}\Pi^{\tau}_{A,k}e^{-i\lambda \ln [p(a^{\textrm{fin}}_{k})]}\otimes\mathbbm{1}_{B}\right)\Phi\left(\sum_{m}\Pi^{\textrm{in}}_{A,m}e^{i\lambda \ln[p(a^{\textrm{in}}_{m})]}p(a_{m}^{\textrm{in}})\otimes\rho_{\textrm{B,in}}\right)\right].\nonumber \\
&&
\end{eqnarray}
Recalling the spectral decompositions of the initial and final density operators, $\rho_{\textrm{A,in}}\equiv\sum_{m}\Pi^{\textrm{in}}_{A,m}p(a^{\textrm{in}}_{m})$ and $\rho_{A,\tau}\equiv\sum_{k}\Pi^{\tau}_{A,k}p(a^{\tau}_{k})$, with eigenvalues $p(a^{\textrm{in}}_{m})$ and $p(a^{\tau}_{k})=p(a^{\textrm{fin}}_{k})$, we get
\begin{equation}
\sum_{k}\Pi^{\tau}_{A,k}e^{-i\lambda \ln[p(a^{\textrm{fin}}_{k})]} = \sum_{k}\Pi^{\tau}_{A,k}e^{-i\lambda \ln[p(a^{\tau}_{k})]} = \sum_{k}\Pi^{\tau}_{A,k}p(a^{\tau}_{k})^{-i\lambda} = \left(\rho_{A,\tau}\right)^{-i\lambda},
\end{equation}
and
\begin{equation}
\sum_{m}\Pi^{\textrm{in}}_{A,m}e^{i\lambda \ln[p(a^{\textrm{in}}_{m})]}p(a_{m}^{\textrm{in}}) =
\sum_{m}\Pi^{\textrm{in}}_{A,m}p(a_{m}^{\textrm{in}})^{1 + i\lambda} = \left(\rho_{A,\textrm{in}}\right)^{1 + i\lambda}.
\end{equation}
If we insert these expressions into Eq.~(\ref{eq:G-appendix}) we obtain the expression for the characteristic function $G_A(\lambda)$ given in Eq.~(\ref{eq:G_A}). Analogously we can derive Eq.~(\ref{eq2}) for $G_B(\lambda)$. In a similar way we can derive the characteristic function $G_{A - B}(\lambda)$ of the stochastic entropy production of the composite system $A - B$:
\begin{equation}
 G_{A - B}(\lambda)=\mathrm{Tr}\left[\rho_{\tau}^{-i\lambda}\Phi(\rho_{\mathrm{in}}^{1+i\lambda})\right]\,.
\end{equation}

\section*{References}

\nocite{*}
\bibliographystyle{iopart-num}
\bibliography{references}

\end{document}